\documentclass{emulateapj}
\usepackage{apjfonts}
\usepackage{natbib}
\usepackage{graphicx}
\usepackage{epstopdf}
\usepackage{multirow} 
\usepackage{longtable,lscape} 
\usepackage{color}
\usepackage{xspace}
\usepackage{xargs}

\newcommand{\M}{$\log M_\star/M_\odot$}
\newcommand{\sex}{{\scshape Sextractor}\xspace}
\newcommand{\dev}{{de Vaucouleurs}\xspace}
\newcommand{\ser}{S\'ersic\xspace}
\newcommand{\dser}{double S\'ersic\xspace}
\newcommand{\nB}{$non-$CG\xspace}
\newcommand{\nBs}{$non-$CGs\xspace}
\newcommand{\B}{CG\xspace}
\newcommand{\Bs}{CGs\xspace}
\newcommand{\eg}{e.g.\xspace}
\newcommand{\nC}{$non-$CG\xspace}
\newcommand{\nCs}{$non-$CGs\xspace}
\newcommand{\C}{CG\xspace}
\newcommand{\Cs}{CGs\xspace}

\shorttitle{Central Group galaxies assembly history}
\shortauthors{Vulcani et al.}

\begin{document}
\title{Understanding the Unique Assembly History of Central Group Galaxies} 

\author{Benedetta Vulcani\altaffilmark{1}}
\author{Kevin Bundy\altaffilmark{1}}
\author{Claire Lackner\altaffilmark{1}}
\author{Alexie Leauthaud\altaffilmark{1}}
\author{Tommaso Treu\altaffilmark{2,3}}
\author{Simona Mei\altaffilmark{4,5}}
\author{Lodovico Coccato\altaffilmark{6}}
\author{Jean Paul Kneib\altaffilmark{7}}
\author{Matthew Auger\altaffilmark{8}}
\author{Carlo Nipoti\altaffilmark{9}}

\affil{\altaffilmark{1}Kavli Institute for the Physics and Mathematics of the Universe (Kavli IPMU, WPI), Todai Institutes for Advanced Study, the University of Tokyo, Kashiwa, 277-8582, Japan}
\affil{\altaffilmark{2}Department of Physics, University of California, Santa Barbara, CA 93106, USA}
\affil{\altaffilmark{3}Division of Astronomy and Astrophysics, University of California, Los Angeles, CA 90095-1547, USA}
\affil{\altaffilmark{4}GEPI, Observatoire de Paris, Section de Meudon, 5 Place J. Janssen, 92190 Meudon Cedex, France}
\affil{\altaffilmark{5}Universit\'e Paris Denis Diderot, 75205 Paris Cedex 13, France}
\affil{\altaffilmark{6}European Southern Observatory, Karl-Schwarzschild-Stra\ss{}e 2, D-85748 Garching bei Muenchen, Germany}
\affil{\altaffilmark{7}Laboratoire d'Astrophysique, Ecole Polytechnique F\'ed\'erale de Lausanne, Observatoire de Sauverny, CH-1290 Versoix, Switzerland}
\affil{\altaffilmark{8}Institute of Astronomy, Madingley Road, Cambridge CB3 0HA, UK}
\affil{\altaffilmark{9}Department of Physics and Astronomy, Bologna University, viale Berti-Pichat 6/2, I-40127 Bologna, Italy}
\email{benedetta.vulcani@impu.jp}

\begin{abstract}
Central Galaxies (CGs) in massive halos  live in unique environments with formation histories  closely linked to that of the host halo. In  local clusters they have 
larger sizes ($R_e$) and lower velocity dispersions ($\sigma$) at fixed stellar mass $M_\ast$, and much larger $R_e$ at a fixed $\sigma$ than field and satellite galaxies ($non-$CGs).  Using spectroscopic observations of group galaxies selected from the COSMOS survey, we compare the dynamical scaling relations of early-type CGs and $non-$CGs at $z\sim$0.6, to distinguish possible  mechanisms that produce the required evolution.
CGs are systematically offset towards larger $R_e$ at fixed $\sigma$ compared to {\it non}-CGs with similar $M_\ast$. The CG $R_e-M_\ast$ relation also shows differences, primarily driven by a sub-population ($\sim$15\%) of galaxies with large $R_e$, while the $M_\ast-\sigma$ relations are indistinguishable.  These results are accentuated when double \ser profiles,  which better fit light in the outer regions of galaxies, are adopted. They suggest that even group-scale CGs can develop extended components by these redshifts that can increase total $R_e$ and $M_\ast$ estimates by factors of $\sim$2.  
To probe the evolutionary link between our sample and cluster CGs 
, we also analyze  
two cluster samples 
at $z\sim0.6$ and $z\sim0$.  We find similar results for the more massive halos at comparable $z$, but much more distinct CG scaling relations at low-z.  Thus, the rapid, late-time accretion of outer components, perhaps via the stripping and accretion of satellites, would appear to be a key feature that distinguishes the evolutionary history of CGs. 
\end{abstract}
\keywords{galaxies: clusters: general -- galaxies: distances and redshifts -- galaxies: evolution -- galaxies: groups: general -- galaxies: dynamics}

\section{Introduction}\label{intro}
Galaxy scaling relations among properties such as mass, size, velocity dispersion, luminosity, and color,  provide valuable insight into galaxy structure and evolution. Early-type (ellipticals and S0s) galaxies, in particular, form a relatively homogeneous population that is described by well-defined scaling relations. The stellar velocity dispersion, $\sigma$, \citep{minkowski62, FJ76} and projected half-light radius, $R_e$, \citep{kormendy77} correlate with galaxy luminosity and stellar mass.  These trends reflect underlying virial relations and are often expressed in terms of the Fundamental Plane (FP) \citep{faber87, dressler87, DD87}.

Reconciling the tightness of the FP with the growth by merger postulated by  hierarchical models
(e.g. \citealt{forbes98}) has 
motivated much effort on understanding how scaling relations are affected by mergers, which predominantly move early-type galaxies {\em along} scaling relations (e.g. \citealt{nipoti03, FM10}). This expectation is in agreement with the modest evolution of the FP since $z\sim 1$, and defined in terms of stellar mass ($M_*$), i.e., removing the effects of the evolution of the stellar population \citep[e.g.,][]{treu05, auger10}.  

Recent work, however, has emphasized the significant evolution observed in projections of the FP that relate size and mass.  Especially remarkable are the compact and massive red ``nuggets'' seen predominantly at $z \sim 2$ that are the expected progenitors of at least some present-day ellipticals \citep{trujillo06,van-dokkum08,bezanson09}.  A number of physical explanations for their significant size growth have been proposed \citep[see,][]{hopkins10a}.  The most popular involve mergers \citep[e.g.,][]{naab09, nipoti12}, especially minor mergers \citep{hopkins10b, trujillo11}, although observations indicate that even the minor merger rate may be insufficient \citep{newman12, cimatti12, sonnenfeld14}.

One way to gain insight into this problem is to study an extreme population, namely the central galaxies (CGs) in massive dark matter halos.  Often referred to as Brightest Cluster Galaxies (BCGs) \citep[e.g.,][]{BG83,JF84}, CGs provide a valuable laboratory because their unique location ties their assembly history to that of the parent halo \citep[e.g.,][]{coziol09}, making them subject to a possible increase in mergers and accretion from tidal stripping events, as well as different gas cooling and heating mechanisms.  Likely as a result, BCGs are offset from early-type scaling relations at the present day, with larger sizes and lower velocity dispersions at fixed luminosity  \citep{thuan81, hoessel87, schombert87, oegerle91, lauer07, liu08, bernardi09}.  Understanding the origin of these offsets, especially the increase in sizes, can provide insight into the processes that drive the more subtle evolution of normal early-type galaxies.

As with the red nuggets mystery, theoretical explanations for the offset scaling relations of BCGs tend to rely on mergers and can be classified in two different categories: 1) merger-driven changes that fundamentally alter the resulting structure of the remnant, and 2) minor-merging induced accretion of low-density stellar material that builds envelopes at large radii.  Arguing in favor of the first explanation, \cite{boylan06} performed a series of major merger simulations and studied the spatial and velocity structure of the remnants.  Regardless of orbital energy or angular momentum, the remnants remained confined to the FP, although their location on projected relations depended on the orbits assumed.  \cite{boylan06} used these simulations to argue that infall along dark matter filaments leads to an increase in radial mergers among CGs that can produce the offsets observed.  When evaluated in a cosmological context, \cite{delucia07} showed that late-time dissipationless merging produces simulated BCGs that show little scatter in luminosity over a wide range of redshifts, as observed (\eg \citealt{sandage72, postman95,aragon93, stanford98}).

In the second category are mechanisms such  tidal stripping of cluster galaxies \citep{gallagher72, richstone75, richstone76,merritt85}, and an uncertain relationship with the formation of an intracluster light (ICL) component (e.g. \citealt{gonzalez05, zibetti05, lauer07}) which may be the result of tidal stripping at very large radii (e.g., \citealt{weil97, puchwein10, rudick11, martel12, cui14} and references therein).  This inside-out growth, with the accumulation of stars in the distant outskirts of CGs, has also been reproduced in hydrodynamical zoom-in simulations  \citep{naab09, feldmann10}. 

Finally, mechanisms like galactic ``cannibalism'' (the merging or capture of cluster satellites due to dynamical friction; \citealt{ostriker75, white76, ostriker77, nipoti04}) might contribute to both the categories. Indeed, the dynamical friction is expected to act more efficiently on more massive systems, while its time scale is long for low mass satellites.

Moreover, existing spectroscopic studies  found that for many CGs the properties of the stellar populations in the outskirts (age, metallicity, and $\alpha$-enhancement) are different from those in central regions \citep{coccato10, greene13, pastorello14}, consistent with the accretion scenario.

In principle, these models could be tested by fitting secondary, extended components to observed light profiles.  Indeed, many studies have highlighted the multi-component nature of early-type profiles  (e.g., \citealt{caon93, lauer95, kormendy99, graham03, ferrarese94, ferrarese06, kormendy09, dullo12, bernardi14}).  Unfortunately, as we emphasize in this work, multi-component fits are often highly degenerate, even when high-resolution and exquisite depths are achieved for local samples (see, e.g., \citealt{huang13}).  It is therefore important to bring to bear additional information encoded in the scaling relations when evaluating these proposed scenarios.

The goal of this paper is to enable such an evaluation by extending \Cs scaling relations to both higher redshifts and lower halo mass than has been previously studied.  We aim to understand which mechanisms are most important and whether there is a critical halo mass or redshift at which \Cs differentiate from the rest of the early-type population.  A key advantage of working at the lower mass scale of galaxy groups is that the  \Cs are more modest in terms of mass and luminosity.  It is therefore easier to find \nCs counterparts with similar mass and morphology, both in the field and in groups, and test whether the unique properties of \Cs are driven by deep-seated structural changes or the accretion of outer components.

By studying a sample in the COSMOS field, we can make use of previous efforts to determine robust group and membership catalogs \citep{leauthaud10, george11}, carefully identify \Cs \citep{george12}, and take advantage of high-resolution imaging from the Hubble Space Telescope (HST) \citep{koekemoer07} and multi-wavelength observations used to derive precise photometric redshifts \citep{ilbert10}.  To this legacy data set, we describe our addition of targeted deep spectroscopy from the Very Large Telescope (VLT) to derive accurate stellar velocity dispersions for both \Cs and \nCs, and present a detailed analysis of profile fitting to the HST imaging.

Throughout this paper, we assume $H_{0}=72 \, \rm km \, s^{-1} \,
Mpc^{-1}$, $\Omega_{m}=0.258$, and $\Omega_{\Lambda} =0.742$ \citep{hinshaw09}.  The \cite{chabrier03} initial mass function (IMF)  in the mass range 0.1--100 $M_{\odot}$ is adopted.

\section{The sample}
\subsection{COSMOS groups}
We use an X-ray-selected sample of galaxy groups from the COSMOS field \citep{scoville07}. 
As presented by \cite{george11, george12}, 
the sample of galaxy groups has been selected from an X-ray mosaic combining images from the XMM-Newton \citep{hasinger07} and Chandra \citep{elvis09} observatories following the procedure of \cite{finoguenov09, finoguenov10}.  Once extended X-ray sources have been detected, a red sequence finder has been  employed on galaxies with a projected distance less than 0.5 Mpc from the centers to identify an optical counterpart and determine the redshift of the group, which has been then refined with spectroscopic redshifts when available.\footnote{
This sample is more complete than comparable group catalogs selected via red-sequence and 3D redshift overdensity methods \citep{wilman05, gerke07} because  X-ray selection better traces halo mass \citep{nagai07} and avoids incompleteness and sparse sampling uncertainties common in spec-z samples.} 

Member galaxies have been selected according to their photometric redshifts and proximity to X-ray centroids. A Bayesian membership probability has been assigned to each galaxy by comparing the photometric redshift probability distribution function to the expected redshift distribution of group and field galaxies near each group. From the list of members, the galaxy with the highest stellar mass within an NFW scale radius of the X-ray centroid is selected as the group center. 
A final membership probability has been assigned by repeating the selection process within a new cylinder recentered on this galaxy.

The COSMOS \Cs we study in this work were originally identified in \cite{george11} who referred to them as  the ``Most Massive Central Galaxies''.  They are defined as the member galaxy with the highest stellar mass within a radius given by the sum of the group's scale radius and the positional uncertainty of the associated X-ray peak.  This definition was studied in detail and determined to be the optimum choice by \cite{george12} who measured the weak gravitational lensing signal around each of the multiple candidate centers (based on luminosity, stellar mass, and proximity to the X-ray center) to find the one which maximized the lensing signal.   The quality of the selection algorithm was further tested with mock catalogs and spectroscopic redshifts, adding robustness to the sample adopted in this paper. 

\cite{george12} consider only groups with a confident spectroscopic association, far from  field edges, not potentially merging groups and  groups with more than four members identified, for a total of 129 groups. However, to increase the statistics in the scaling relations, we consider all 169 groups with an identified \B,\footnote{Including the groups excluded by \cite{george12}  does not bias results.}  ranging from redshift $0 < z < 1$ and from halo masses $\sim 10^{13} <M_{200c}/M_\odot <10^{14}$, as estimated with weak lensing \citep{leauthaud10}. $M_{200c} = 200\rho_c(4\pi/3)R^3$ is the mass enclosed within  
$R_{200c}$, which is the radius within which the mean mass density equals 200 times the critical density of the universe at the halo
redshift, $\rho_c(z)$. 

\subsection{Extant Data Products}\label{add_data}

We exploit additional COSMOS data, briefly summarized here.  The $HST/ACS$  FW814 imaging is described in \cite{scoville07} and \cite{koekemoer07}, and is used for 
the profile fitting in this work. 

Stellar masses  have been determined by fitting stellar population synthesis models to the spectral energy distributions of galaxies, varying the age, amount of dust extinction, and metallicity in the models. We use the stellar masses from \cite{bundy06, bundy10}, which are based on fits to \cite{bc03} models  and a \cite{chabrier03} IMF, in the mass range 0.1--100 $M_{\odot}$.

To separate passive from star forming galaxies, we  use a  color-color diagrams that include rest-frame UV, optical, and near-IR colors, as presented in \cite{bundy10}. 
Passive galaxies must satisfy the following cuts:
$$ NUV - R > 4.2(1 + z)-0.43 - 0.2(M_K + 20)$$ 
and
$$NUV - R > C_1(z) + C_2(z)(R - J)$$
where  $M_K$ is the rest-frame absolute $K_s$-band magnitude and  the constants, $C_1$ and $C_2$, have been chosen by inspection in redshift bins. For our sample, for z = [0.30, 0.50, 0.70, 0.85] $C_1(z)$ = [4.4, 4.2, 4.0, 3.9] and $C_2(z)$ = [2.41, 2.41, 2.5, 2.6].

\subsection{FORS2 observations and reductions}

Despite the array of data sets available in COSMOS, followup spectroscopy was required to determine accurate velocity dispersions for group \Cs as well as an appropriate control sample of  \nCs.  To accomplish this, a four-night program\footnote{The ESO program ID was 084.B-0523(A).} (PI: S.~Mei) using the FOcal Reducer and low dispersion Spectrograph \citep[FORS2,][]{appenzeller98} on the VLT was executed from 14--17 Februrary 2010.  The holographic 600 RI+19 grism (with filter GG435) was used with 0\farcs6 width slits to acheive a wavelength range of 5900--8000 \AA\ with an instrumental dispersion of $\sigma_{\rm sp} \approx 75$ km s$^{-1}$.  A total of 27 slit plates, each with a field of view of 6\farcm8 by 5\farcm7 and positioned to span roughly 3 COSMOS groups, were observed in 1 hr blocks (45 min on-sky integration time).  A total of 353 targets (85 \Cs and 268 \nCs) were observed.

Slits were allocated with the highest priority to candidate \Cs using a preliminary \C catalog provided by M.~George (private communication).  The second most likely \C candidates were also targeted. The control sample of \nCs was required to have $M_* > 10^{9.5} M_\odot$ and divided into group members and field galaxies.  Group \nCs had to satisfy the membership criteria presented in \cite{george11}.  Field \nCs were chosen from galaxies not associated with any group and additionally were prioritized to match as closely as possible the $M_*$ distribution of \Cs.  No morphology cuts were implemented in the selection because the full range of the morphological distribution of \Cs is relatively broad (although some cuts were later applied in our analysis (see \S\ref{final}).  To obtain a velocity dispersion, targets were required to have a F814W $I$-band {\sc mag\_auto} AB magnitude brighter than $20.5$ and a redshift (either spectroscopic or photometric) in the range $0.2 < z < 0.9$.  When room on the slit masks was available, additional sources related to other science goals were also targeted.

For the present work, reductions were performed using scripts from the Carnegie Python Distribution\footnote{{\tt http://code.obs.carnegiescience.edu/\\carnegie-python-distribution}} (CarPy) originally packaged to reduce spectroscopic observations from LDSS2.\footnote{see {\tt http://astro.dur.ac.uk/\~ams/dan}}  The scripts work primarily on the 2D spectral images, applying bias subtraction, slit tracing, flat fielding, and wavelength rectification based on supplied arc and flat frames.  Traces from multiple exposures are combined and extracted into 1D spectra.  A range of tests and optimizations of the scripts was explored often with fine-tuning required for the wavelength rectification and sky subtraction of individual slits.  

\subsection{Derived quantities}

\subsubsection{Stellar velocity dispersion fitting}\label{sec:vd}

Following \cite{suyu10} and \cite{harris12}, we use a Python-based implementation of the velocity dispersion code from \cite{vdmarel94}, expanded to use a linear combination of template spectra (written by M.~W.~Auger).  The code simultaneously fits a linear combination of broadened stellar templates and a polynomial continuum to the data using a Markov Chain Monte Carlo (MCMC) routine to find the probability distribution function of the velocity dispersion ($\sigma$) and velocity ($v$).  We use a set of templates from the INDO-US stellar library containing spectra for a set of seven K and G giants with a variety of temperatures and spectra for an F2 and an A0 giant.  Before fitting, template spectra are convolved to the instrumental resolution determined from the science spectrum. Our reported measurement of $\sigma$ and $v$ are the median of the MCMC distribution, and measurement errors are the semi-difference of the values at the 16 and 84 percentile values.\footnote{We do not fit higher order moments ($h_3$ and $h_4$). However, they would have little effect on the high $\sigma$ values of our galaxies.} Systematic errors due to template variations are accounted for in our fitting routine since the template weights are fitted simultaneously with the velocity dispersion and marginalized over. 

For all the measurements, we use the entire wavelength range to perform the fit,\footnote{We tested that measurements obtained using only specific regions of the spectra gives results that are fairly in agreement with those obtained adopting the entire spectrum.} defining regions to mask to ensure that we fit only the stellar contributions to each spectrum and using a 5th-order polynomial to fit the continuum.  We then exclude from the fit the following emission lines: OII $\lambda$3715-3740, $H_\delta$ $\lambda$4060- 4110, $H_\gamma$ $\lambda$4330-4345, $H_\beta$ $\lambda$4850- 4870, OIIIa $\lambda$4950- 4965, OIIIb $\lambda$5000- 5015, $H_\alpha$ $\lambda$ 6550, 6573. In addition, we mask the main telluric bands and the sky lines.  Every spectrum has been inspected by eye by one of us (B.V.) to mask possible additional bad regions of the spectra that could alter the fit.

\begin{figure}
\begin{center}
\includegraphics[scale=0.4]{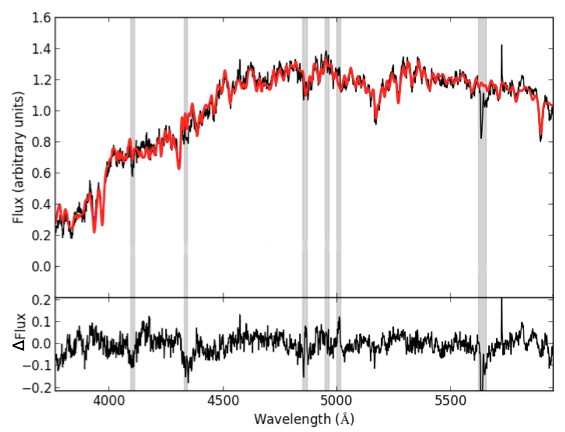}
\caption{Example of a \B spectrum (black line) with a model generated from all nine INDO-US templates and a fifth-order continuum over plotted (red line). The gray shaded areas are the regions not included in the fit, and the lower panel shows the fit residuals. The S/N obtained from the FeMG feature is 31. 
\label{es}}
\end{center}
\end{figure}

Each of these parameters (wavelength range, mask, polynomial order) has been determined by visual inspection of each spectrum with a S/N $\geq$7 pixel $^{-1}$, while galaxies with S/N$<$7 pixel $^{-1}$ have been disregarded from the analysis.   Figure \ref{es} shows an example of a typical spectrum, with a model over plotted and the regions masked.  Following \cite{jorgensen95} and taking into account that our  slits have a rectangular shape of $0.6^{\prime\prime}\times R_e$, we correct velocity dispersions to standard central velocity dispersions within 1/8 $R_e$, applying the formula: 
$\sigma=\sigma_{ap}\times \left [ \frac{1.025(0.6\times R_e)^{1/2}}{R_e/8} \right ]^{0.04}$. Such correction produces an increase in the reported central velocity dispersions of 5--10\%.

\subsubsection{Profile fits and size estimates}

Because our ultimate goal is distinguishing processes that affect the fundamental internal structure of galaxies (e.g., radial major mergers) from those that affect only the outskirts and leave the core unchanged (e.g., growth of outer envelopes), fitting model profiles to HST images of our sample and interpreting the resulting half-light radii estimates is critically important.  As discussed in \S 1, there are several lines of evidence that massive early-type galaxies are made up of multiple components possibly formed at different epochs (e.g., \citealt{gonzalez05}; Hopkins {\it et al.} 2009b; Hopkins {\it et al.} 2009c;  \citealt{dhar10,   dullo13, huang13, bernardi14}).  Ideally, the profile fits to different components would correspond to meaningful information about truly distinct {\em physical} components, their shapes, sizes, and associated masses.  Unfortunately, as we show below, even with the  HST data available for our sample, the best-fit parameters for multi-component fits can be highly degenerate, rendering physical insight from multi-component fits very uncertain.

The first is the widely-used \dev profile. Not only does it facilitate comparisons between our work and others, it also maintains a fixed shape, making comparisons between galaxies easier to interpret. The \dev profile is primarily sensitive to light in the central regions and less sensitive to ``extra'' material in the outskirts. This, in turn, makes the \dev half-light radii systematically small. The second model is a double \ser. Its several extra degrees of freedom ensure that the more complex profiles resulting from significant amounts of light in the galaxy's outskirts can be adequately accounted for, while still accurately modeling the galaxy's central regions. While the half-light radius of either one of the two \ser  components may not be physically meaningful, the total half-light radius is likely to be more sensitive to outer envelopes than the more restricted de Vaucouleurs profile, whose shape cannot adjust to accommodate light at distant radii (e.g., \citealt{gonzalez05, bernardi13}).

We can demonstrate the need for caution when interpreting the half-light radii of the individual \ser components by fitting a different two-component model to the same galaxies and comparing the component half-light radii. We find that fitting a more restrictive double \dev profile to the galaxies is often equivalent to fitting a double \ser profile. For two thirds of both \Cs and \nCs, a double \dev and double \ser profile are indistinguishable based on the reduced $\chi^2$ values, accounting for the extra degrees of freedom in the double \ser profile. This should be contrasted to the difference in $\chi^2$ values between single- and double-component models, which justifies using a double-component model for 90\% of the galaxies in the sample. The two-component \ser and \dev models yield dramatically different half-light radii for the separate components of the galaxies, with scatter on the order of 100\% in the half-light radius of the outer component, while the total half-light radii vary by $\sim$20\%. Thus, the total half-light radii are much more robust. 

The lack of a difference between the goodness-of-fit of double \dev and double \ser profiles also demonstrates that using more complicated models is not necessary for most of the galaxies in this sample, as extra parameters are not justified.  For the minority of galaxies with additional features in the radial profile, more complicated models may be merited \citep{huang13}. In this work, we use the double \ser fits as visual inspection has shown fewer catastrophic failures for the more flexible profile. 

\begin{figure}
\begin{center}
\includegraphics[scale=0.38]{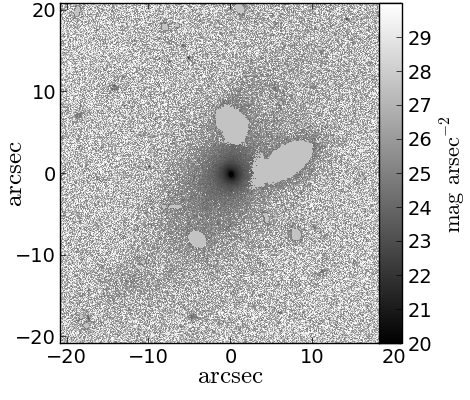}
\includegraphics[scale=0.38]{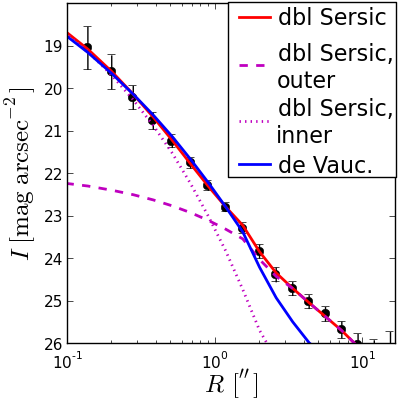}\\
\includegraphics[scale=0.38]{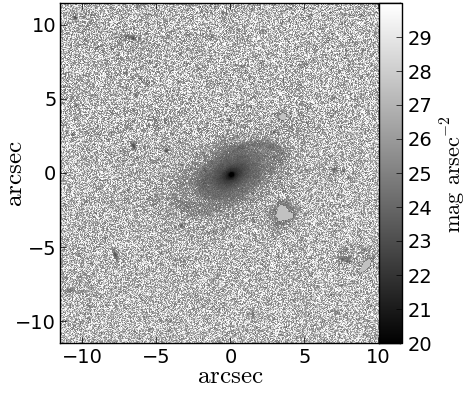}
\includegraphics[scale=0.38]{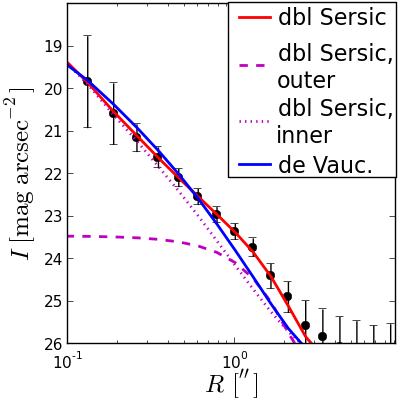}
\caption{Example of a \B whose outer \ser component is representative of an outer envelope ({\it top}), and of a \nC whose two \ser components are attributed to a bulge and disk ({\it bottom}). Both images are are arcsinh stretched.
\label{profile}}
\end{center}
\end{figure}

The uncertainty in interpreting the half-light radii of the individual \ser components can also by illustrated by looking at specific galaxies in the sample. For \Cs with significant outer envelopes, the two \ser components often 
align well with the inner galaxy and outer envelope. For smaller systems, the two \ser components are often attributed to a disk and a bulge, or a bright central region and the bulk of the galaxy.  
Figure \ref{profile} shows two galaxies for which the single \dev does not provide an adequate fit, but the added \ser profile fits different physical structures in the galaxy. The top panels show a \C in which the larger \ser component refers to the outer envelope, while the single \dev profile fails to capture the outer envelope. The lower panels in Figure \ref{profile} show a \nC. In this case, the larger \ser profile models the disk of the galaxy, not an outer envelope. Whether or not the outer \ser component refers to an outer envelope depends on the galaxy morphology, particularly the brightest features in the galaxy. This ambiguity further justifies using the \emph{total} half-light radius instead of the radii for the individual components.

With these choices defined, we fit both \dev and double \ser models to the {\it HST}/ACS FW814 images of both \Cs and \nCs in our sample.  These images have been drizzled such that they have a pixel scale of $0.03"/\mathrm{pix}$ \citep{koekemoer07}. For each galaxy, we create a postage stamp image with a size $4\times a_{iso}$, where $a_{iso}$ is the major axis diameter of the \sex isophotal area. We also mask any other sources from by \sex in the postage stamp image. For 16 of the images, we manually mask nearby sources which would otherwise disrupt the model fitting. 

Each model component is described by seven parameters: flux normalization, half-light radius, \ser index (fixed to $n=4$ for \dev profiles), axis ratio, centroid position, and position angle. In the two-component \ser model, the centroid positions of both components are held to the same value, but all other parameters are free and independent. The models are fit using a version of the galaxy fitter used by \cite{lackner12}, modified to accommodate {\it HST} images. Briefly, we obtain the best-fitting model parameters by performing a $\chi^2$ minimization over the difference between the PSF-convolved model and the galaxy image, weighted by the measured inverse variance of the image. Although the models used here are symmetric under rotations, we do not bin pixels in radius, but perform a full two-dimensional fit.  The initial conditions for the model fits are taken from single-component \ser fits, which are not used in the subsequent analysis. The initial conditions for the half-light radius and axis ratio for the single-component fits are derived from the \sex isophotal area. 

The total half-light radius of the double \ser fits is computed numerically by determining the size of the ellipse that contains half the flux.  For both the \dev and double \ser profiles, we report half-light radii along the major-axis.  The axis ratio and the position angle of the ellipse are taken from the single-component S\'ersic fit.  While the least-squared fitting does report errors on the half-light radii, these are likely under-estimated. Instead, we compute the scatter in the size measured using different profiles (\ser, \dev, double \ser, double \dev, \dev$+$\ser, exponential$+$\ser, \dev$+$ exponential). The typical scatter in these half-light radii is $\sim20\%$, and we use this as the uncertainty for all half-light radii, from now on called simply sizes for brevity.

\subsubsection{Stellar mass corrections}\label{mascor}

While ideally independent, to derive the properties compared in the usual galaxy scaling relations requires similar assumptions.  In this section we make an attempt to reduce the systematics due to the different assumptions made to estimate sizes and masses. Indeed, sizes have been determined through a profile fit to the galaxy's surface brightness, while masses were estimated under a different set of assumptions for the shape of the same surface brightness profile. In particular,  
stellar masses  have been scaled to a Kron ``total magnitude'' ({\sc mag\_auto}) as measured in the $K$-band, not to a luminosity determined from a multi-component profile fit \citep{bundy10}. 

To obtain mass estimates that are more closely associated with the profiles we use to derive size estimates, we apply the following correction.  We first assume that the $K$-band {\sc mag\_auto} 
 is obtained with the same SExtractor photometric estimator as the F814W {\sc mag\_auto}  magnitudes
\citep{leauthaud10}, and ignore potential color gradients.
We then compute the difference in magnitude, $\delta m$, between the F814W {\sc mag\_auto} and the magnitude associated with either the \dev or double \ser fit described above.  Assuming the same $M/L$ for all early-types, the resulting difference in log mass, $\delta \log M_*$, is given by $\delta \log M_* = -\delta m / 2.5$ in units of dex.  We apply the appropriate correction to $M_*$ for either the \dev or double \ser fits as required.  In parallel with the discussion on size estimates above, the resulting $M_*$ values referenced to the \dev profile are dominated by a galaxy's central region and are less sensitive to mass in the outskirts.  The $M_*$ values referenced to the double \ser profile, instead,  are a more accurate representation of the {\em total} mass profile, including material potentially in an outer envelope.

\begin{figure}
\centering
\includegraphics[scale=0.3]{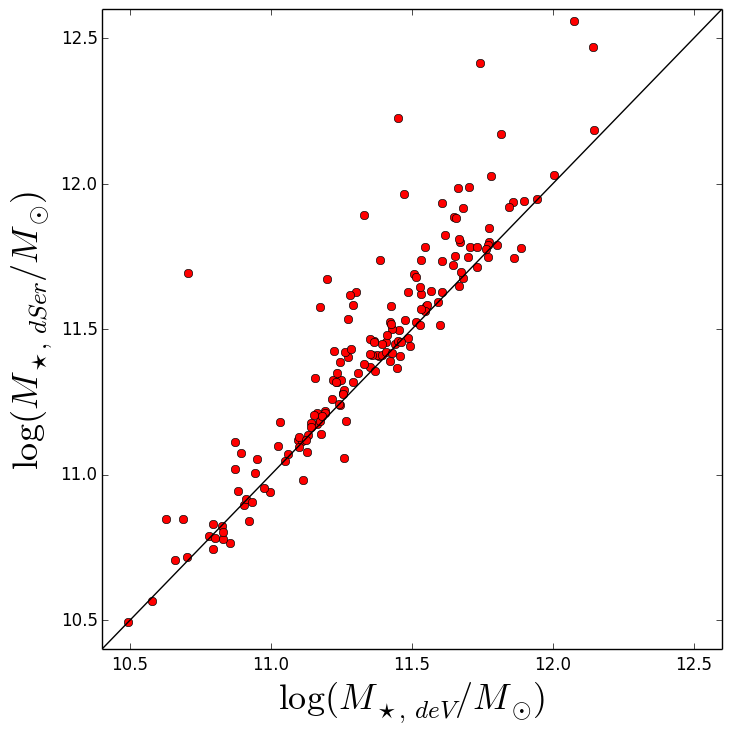}
\caption{Comparison of the stellar mass estimates associated with the two profiles we adopt to derive size estimates. Black line represents the identity. 
\label{cfrmass}}
\end{figure}

Figure \ref{cfrmass} shows the comparison between the stellar masses corrected using the two different profiles. At low masses, estimates are in agreement, while for \M$>$11 double \ser masses are systematically larger than  \dev masses, indicating that the former profile is likely including an outer envelope for massive galaxies.

\subsection{The final sample properties} \label{final}
Table \ref{inventory} describes our final sample, reporting the number of galaxies removed for various reasons at each stage of analysis. We note that, with respect to the original sample,  we have applied an additional cut in \ser index ($n>2.5$), to remove those galaxies that are characterized by a non negligible disk, hence for which a \dev profile is not accurate. 

On the whole, our final sample consists of 107 \nBs and 41 \Bs for which a robust $\sigma$ is available. In addition, when possible, we enlarge our \B sample by adding 128 \Bs without $\sigma$ ($all-$CGs), having checked that the mass distribution is comparable for the two samples. 

\renewcommand{\tabcolsep}{3pt}
\begin{table}
\centering
\caption{The spectroscopic sample at various stages of our data analysis}
\begin{tabular}{lcc}
\hline
{\bf Step} 													& {\bf \Bs} 	&{\bf \nBs} \\ 
Observed with original target definition 			& 	85 &  	268 \\ 
Catastrophically bad spectroscopy                             & -12 &-28 \\
Untrustworthy velocity dispersion (S/N$<$7)							& -25 & 	-115 \\
Spectroscopically confused pairs							& -4	& -0 \\
Catastrophic failure in profile fitting and size determination  &	-2 & 	-1 \\
\ser index $n < 2.5$												&	-3	&	-17	\\
\hline
Fnal Sample													&	41	& 107 \\
\hline
\hline
\end{tabular}
\label{inventory}
\end{table}

\begin{figure}
\begin{center}
\includegraphics[scale=0.4]{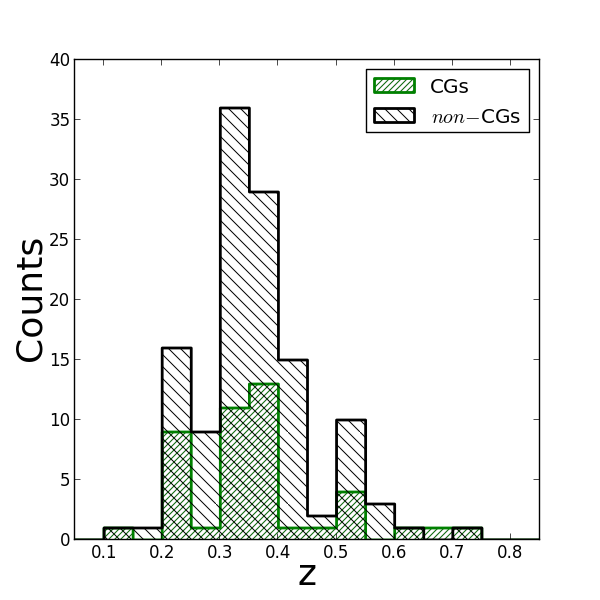}
\caption{Redshift distribution for all galaxies (\Bs: green  line, \nBs: black line) with a robust velocity dispersion measurement   (see \S\ref{sec:vd} for details). }
\label{z}
\end{center}
\end{figure}

Figure \ref{z} shows the redshift distribution of the spectroscopic sample, for $all-$\Bs and \nBs separately.  The spanned redshift range goes from $z$$\sim$0.1 to 0.8, even though in both samples most of the galaxies are located between $z=$0.2-0.6. A Kolmogorov-Smirnov (K-S) test, which quantifies the probability  that two data sets are drawn from the same parent distribution, can not exclude the hypothesis that distributions are similar.

\section{Additional samples from the literature}\label{other_sample}

To put our COSMOS results in the context of previous work on clusters at both high and low redshift, we 
make specific use of additional samples from the literature.  Taken together, we warn that these samples are inhomogeneous in both data and methods and use different definitions of CGs.  We do not attempt quantitative intra-sample comparisons, but relative comparisons within each sample between CGs and {\it non-}CGs are robust.  In all cases where relevant, we have converted derived quantities to our adopted cosmology and stellar masses to a \cite{chabrier03} IMF.

\subsection{Cluster at intermediate z: EDisCS}
To compare our CG sample in COSMOS groups to CGs in more massive clusters and groups at similar redshifts, we use the ESO Distant Cluster Survey (EDisCS - \citealt{white05}),
a catalog of 25 clusters and groups. 
EDisCS provides information of galaxies both in clusters (both central and satellites) and in the field. 
Clusters and groups have redshifts between 0.4 and 0.9 and structure velocity dispersions between 166 and 1080 $km \, s^{-1}$ \citep{halliday04, clowe06, milvang08}, yielding estimated halo masses between  $10^{12}$ and $1.5\times 10^{15} M_\odot$, with a mean value of $\sim$4$\times$10$^{14} M_\odot$. Photo-$z$ membership  was established using a modified version of the technique developed in \cite{brunner00} \citep{delucia04, delucia07_bis, pello09}. Such an approach rejects spectroscopic non-members while retaining at least 90\% of the confirmed cluster members. A posteriori, it was verified that above \M=$\sim$10.2, $\sim$20\% of the galaxies classified as photo-$z$ cluster members were actually interlopers, and, conversely, only $\sim$6\% of those galaxies classified as spectroscopic cluster members were rejected by the photo-$z$ technique \citep{vulcani11}.  The identification of the CG was based on available spectroscopy, the brightness, color, and spatial distributions of galaxies, their photometric redshifts and  weak lensing maps \citep{white05, whiley08}.

We specifically use the EDisCS data set presented by \cite{saglia10}. Velocity dispersions were measured for all galaxy spectra using the IDL routine pPXF \citep{cappellari04} and corrected using the recipe by \cite{jorgensen95}. 
The half-light radii  were derived by fitting either HST ACS images \citep{desai07} or I-band VLT images \citep{white05} using the GIM2D software \citep{simard02}.
Given the fact that most of the EDisCS HST observations on which the profile fitting was performed employed a strategy similar to that in COSMOS (same camera and similar exposure time), we expect that the surface brightness depth of these images is similar to COSMOS. 
Two-component, two-dimensional fits were performed, adopting a \dev bulge plus an exponential disk convolved with the PSF of the images (for details, see \citealt{simard09}). Unlike \cite{saglia10}, we do not use circularized half-luminosity radii, reporting instead major-axis scale lengths as we adopt for the COSMOS sample.
Rest-frame absolute photometry were derived from SED fitting \citep{rudnick09} and used to derive stellar masses, which were computed adopting the calibrations of \cite{bdj01}. 
and B-V colors, and renormalized using the corrections for an elliptical galaxy given in \cite{djb07}.  For the EDisCS clusters we do not have detailed information on the profile fitting, so we cannot reference $M_*$ values to specific profile choices, as we did for our COSMOS data, therefore we use the original stellar masses as provided by \cite{saglia10}. 

\begin{figure*}
\centering
\includegraphics[scale=0.3]{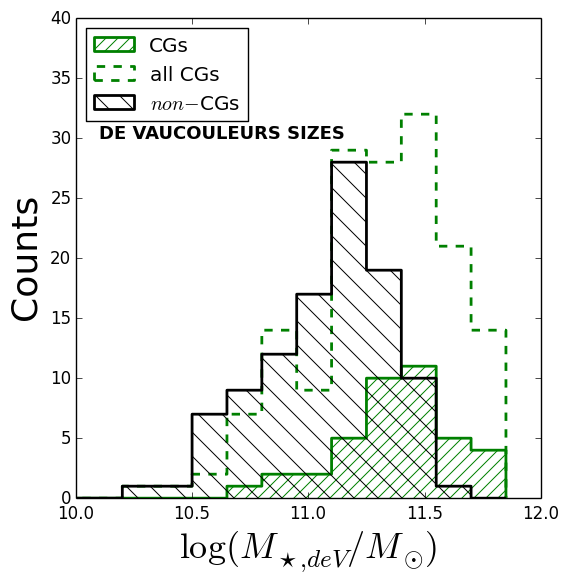}
\includegraphics[scale=0.3]{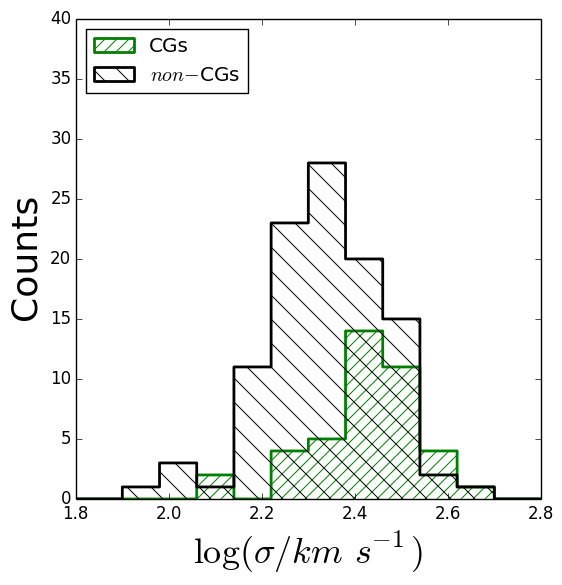}
\includegraphics[scale=0.3]{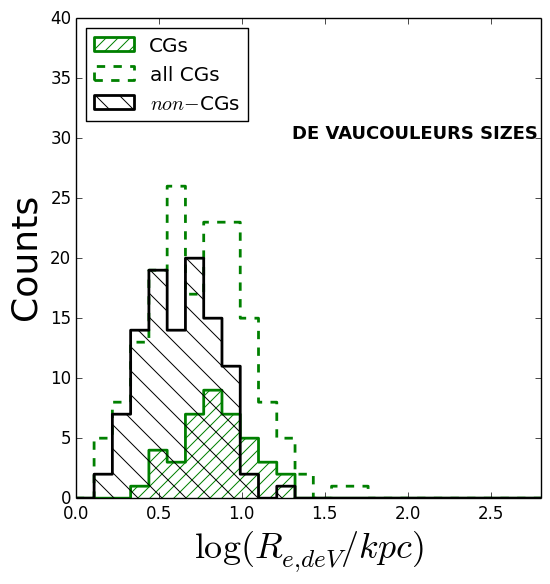}
\includegraphics[scale=0.3]{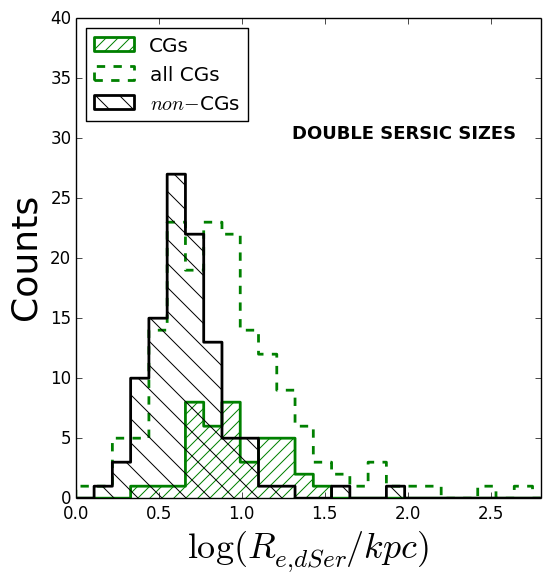}
\caption{\dev stellar mass, velocity dispersion and effective radius distributions for \Bs (green lines) and \nBs (black lines). In the mass and size distribution panels we show the \B distribution both for the entire sample (dashed line) and for the subsample with a measured velocity dispersion (solid line). 
\label{distr}}
\end{figure*}

\subsection{Cluster at low z: WINGS}
For a low-$z$ comparison that allows for an evaluation of potential redshift evolution, we take advantage of the WIde-field Nearby Galaxy-cluster Survey (WINGS- \citealt{fasano06}), which gives us information on 77 galaxy clusters at $0.04<z<0.07$ that span a halo mass range of $\sim 1.5\times10^{14}-2.1\times 10^{15} M_\odot$. Integrated and aperture photometry have been obtained for galaxies using \sex \citep{bertin96}.  For CGs, great attention has been paid to avoid mutual photometric contamination between big galaxies with extended stellar haloes and smaller, halo-embedded companions \citep{varela09}. Briefly, the largest galaxies in each cluster were carefully modeled with IRAF-ELLIPSE and removed from the original images in order to allow a reliable masking of the small companions when performing the surface photometry of the big galaxies themselves.  The surface brightness limit was computed setting the detection threshold to 4.5$\sigma_{bg}/arcsec^2$, with $\sigma_{bg}$ the standard deviation of the background signal.  This limit translates to a detection limit of $\mu_{Threshold}(V) \sim 25.7 mag/arcsec^2$.  The  WINGS surface photometry has been obtained using GASPHOT \citep{pignatelli06, donofrio14}) and non-circularized, major-axis sizes have been estimated using a single \ser law.

Galaxy stellar mass estimates were derived using the \cite{bdj01} relation which correlates the stellar mass-to-light ratio with the optical colors of the integrated stellar population \citep{vulcani11}.  Using the magnitudes computed by the GASPHOT routine, we apply corrections to the stellar masses given in \cite{vulcani11} similar to those for the COSMOS galaxies, to reference the $M_*$ estimates to adopted single-\ser surface brightness profiles.  The single \ser profile is intermediate between the \dev and double \ser profiles in its sensitivity to light at large radii. 

Here we use the CG sample presented in \cite{fasano10}, which consists of 75 galaxies. However, only 55 of these have a reliable estimates of the velocity dispersions from the published data of the NFPS and SDSS-DR6 surveys \citep{bernardi03, smith04}.  

\section{Results}\label{res}
In this section, we present a comparison of the dynamical and structural properties of \Cs and \nCs drawn from the $z\sim 0.6$ COSMOS group sample.  Our goal is to use differences between these populations and their scaling relations to place constraints on the physical mechanisms that drive offsets in the properties of \Bs at low-$z$ and can account for the evolutionary paths of early-type galaxies more generally.  

As discussed in \S\ref{intro}, we seek observational signatures that can distinguish between structural evolution occurring in the cores of galaxies from growth occurring primarily at large radii.  Unfortunately, our observations do not offer a robust way to cleanly separate the inner and outer components of galaxies in our sample.  Instead, we make use of different combinations of observables which we argue are sensitive to properties of either the inner or outer regions.  As an example, we investigate size and $M_*$ estimates using fits to two types of model surface brightness profiles: the single \dev and the double \ser. The former adequately models the inner, "primary" region of early-type galaxies, while the latter is also sensitive to light in the outer regions.

We start by characterizing \Cs and \nCs in terms of  their global distributions in stellar mass, velocity dispersion and size distributions. Figure \ref{distr} shows that ($all-$)\Cs are  larger in all quantities than \nCs and  the \dser sizes are larger on average than the \dev sizes. The spanned  $\sigma$ range is comparable in the two samples. 
Both considering the  whole sample and the sample fro which a robust $\sigma$ is available, the K-S test excludes the hypothesis that the two distributions are similar for each quantity at $>99.9\%$ level. 

The observed overlap between the COSMOS-group \C and \nC stellar mass distributions indicates that, as opposed to massive clusters where the CGs are by far the most massive objects and it is very difficult to find equally massive \nCs for comparison, we can begin to disentangle the influence of the mass from the influence of the environment (central/satellite).
Nonetheless, \Cs and \nCs still show different $M_*$ distributions that are not perfectly matched (\Cs are on average larger).  To make further progress and determine whether  \Cs are simply  the tail of the general population, in the next section, we investigate the scaling relations of these two populations.

\renewcommand{\tabcolsep}{3pt}
\begin{table*}
\centering
\caption{Scaling relations}
\begin{tabular}{llcccccc}
\hline
\hline
\multirow{2}{*}{Y} & \multirow{2}{*}{X}& \multirow{2}{*}{sample} & \multicolumn{3}{c}{free parameters} & \multicolumn{2}{c}{fixed slope}\\
				&						&						&Slope $a$  & Intercept $b$&Scatter&Slope $a$  & Intercept $b$\\ 
\hline
\multirow{2}{*}{$\sigma$} & \multirow{2}{*}{$M_{\star, \, deV}$} & \Bs &0.28$\pm$0.07&-0.7$\pm$0.8 & 0.06$\pm$0.02 &0.26 & -0.51$\pm$0.01\\
						&							&	\nBs &0.26$\pm$0.04&-0.5$\pm$0.5& 0.07$\pm$0.02 & 0.26 & -0.51$\pm$0.09\\
\multirow{2}{*}{$R_{e, deV}$} & \multirow{2}{*}{$M_{\star, \, deV}$ }& $all-$\Bs &0.6$\pm$0.1&-6$\pm$1 & 0.04$\pm$0.01 &0.55 & -5.46$\pm$0.03\\
						&							&	\nBs &0.55$\pm$0.07&-5.5$\pm$0.8 & 0.023$\pm$0.005&0.55 & -5.49$\pm$0.02\\
\multirow{2}{*}{$R_{e, dSer}$} & \multirow{2}{*}{$M_\star, \, dSer$} & $all-$\Bs & 0.7$\pm$0.1 &-7$\pm$1 &0.1$\pm$0.1 &0.51 & -4.85$\pm$0.03 \\
						&							&	\nBs &0.51$\pm$0.09 &-5$\pm$1 &0.07$\pm$0.03& 0.51 &-4.94$\pm$0.02\\
\multirow{2}{*}{$R_{e, dSer}$} & \multirow{2}{*}{$M_{\star, \, deV}$} & $all-$\Bs &0.6$\pm$0.1&-6$\pm$1& 0.16$\pm$0.02 &0.65 & -6.45$\pm$0.03\\
						&							&	\nBs &0.65$\pm$0.07&-6.5 $\pm$0.8 & 0.025$\pm$0.005 &0.65 & -6.51$\pm$0.02\\
\multirow{2}{*}{$R_{e, deV}$} & \multirow{2}{*}{$\sigma$} & \Bs &0.2$\pm$0.4&0$\pm$1 & 0.05$\pm$0.01 &0.05 & 0.70$\pm$0.03\\
						&							&	\nBs &0.1$\pm$0.2 &0.5$\pm$0.5 &0.04$\pm$0.01 & 0.05& 0.50$\pm$0.02\\
\multirow{2}{*}{$R_{e, dSer}$} & \multirow{2}{*}{$\sigma$} & \Bs &0.6$\pm$0.4&-1$\pm$1 &0.06$\pm$0.02& 0.43 & -0.11$\pm$0.04\\
						&							&	\nBs &0.4$\pm$0.3 &-0.4$\pm$0.6 &0.06$\pm$0.01&0.43& -0.34$\pm$0.02\\
\hline
\end{tabular}
\tablecomments{Fits are of the form $\log Y=a\log X+ b$ with $M_\star$ 
in units of  $M_\odot$, $\sigma$ in units of $km\, s^{-1}$, $R_e$ in units of kpc. 
\label{tab_fits}}
\end{table*}

\subsection{Scaling relations}\label{relations}
\subsubsection{ Size-Stellar mass relation}
\begin{figure*}
\centering
\includegraphics[scale=0.45]{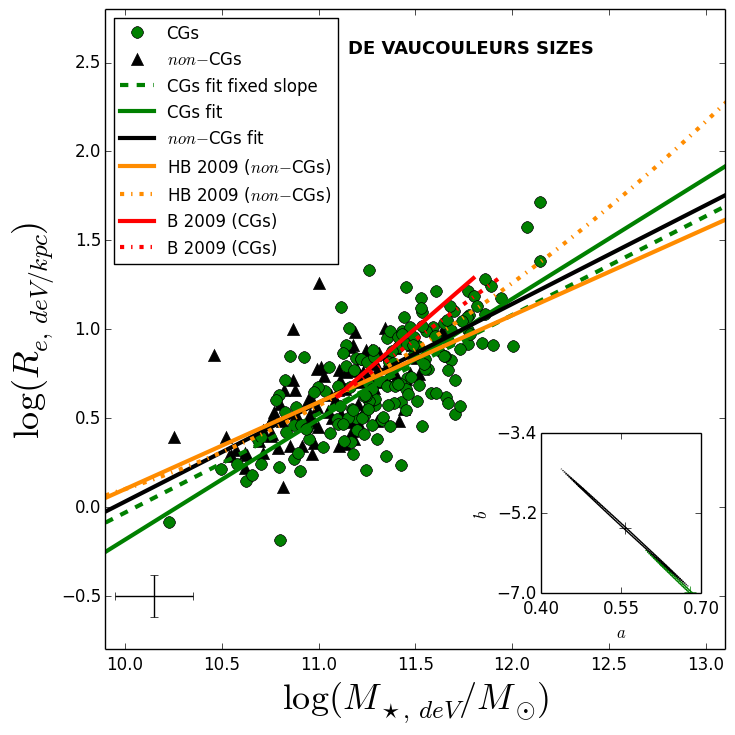}\\
\includegraphics[scale=0.45]{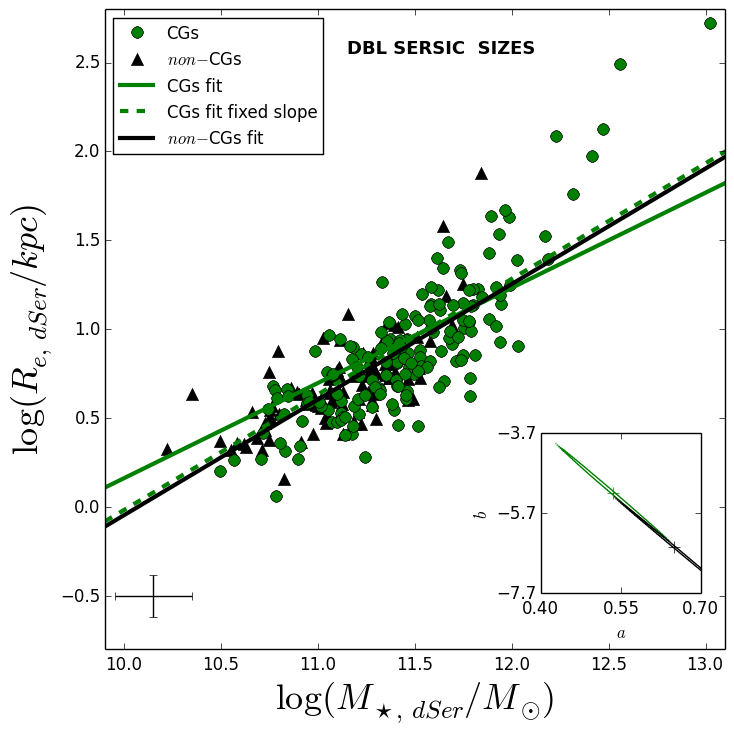}
\includegraphics[scale=0.45]{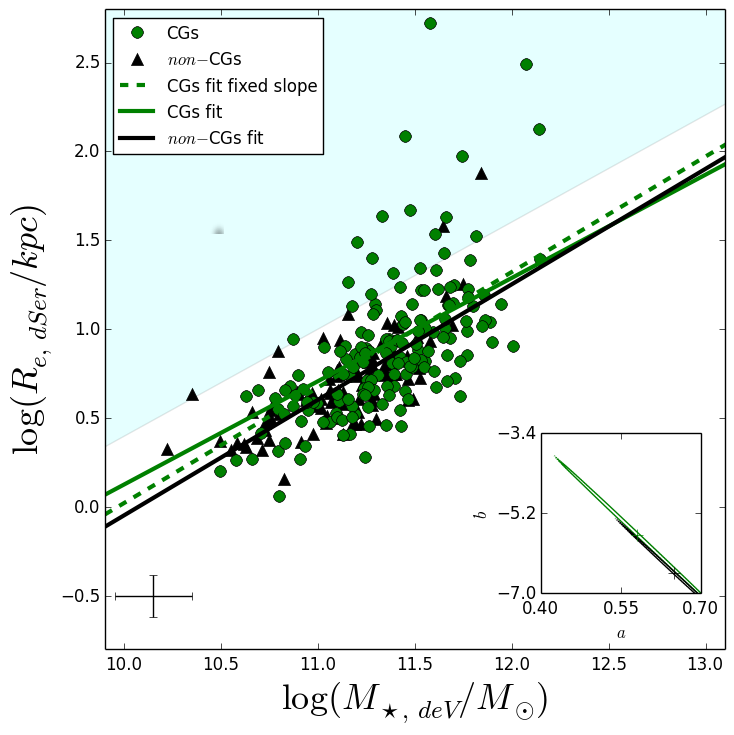}
\caption{$R_e- M_\star$ relation for \Bs (green circles) and \nBs (black triangles) for the entire COSMOS sample. Solid lines are the linear fit to the relations, the green dashed line is the fit to the \Bs when their slope is fixed to be the same as \nBs. In the upper panel sizes have been measured adopting a \dev profile, in the bottom panels with a \dser profile (left: mass correction using the double \ser profile, right: mass correction using the \dev profile). The typical errors are shown in the bottom left corner of each panel. In the inset, the 1-$\sigma$ contour errors to the fits are given.  In the upper panel, the orange lines represent the linear (solid) and quadratic (dotted) fit as given in \citet{hb09} for a low-$z$ sample of early type galaxies, the red lines represent the fits as given in \citet{bernardi09} for CGs in the local universe, for two different CG samples, all fitted with a \dev profiles. In the bottom right panel,  regions at $>1\sigma$ from the \B+\nB relation have been marked in cyan (see text for details). 
\label{re_mass}}
\end{figure*}

\begin{figure}
\centering
\includegraphics[scale=0.45]{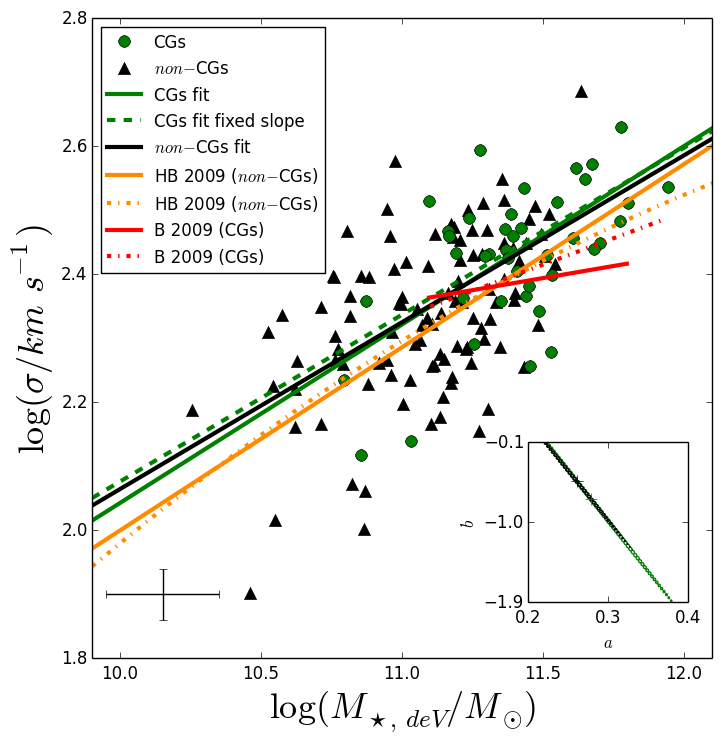}
\caption{$\sigma- M_\star$ relation for \Bs (green circles) and \nBs (black triangles) for the spectroscopic COSMOS sample. Lines and symbols are as in the upper panel of Fig. \ref{re_mass}. 
\label{sigma_mass}}
\end{figure}

\begin{figure*}
\centering
\includegraphics[scale=0.45]{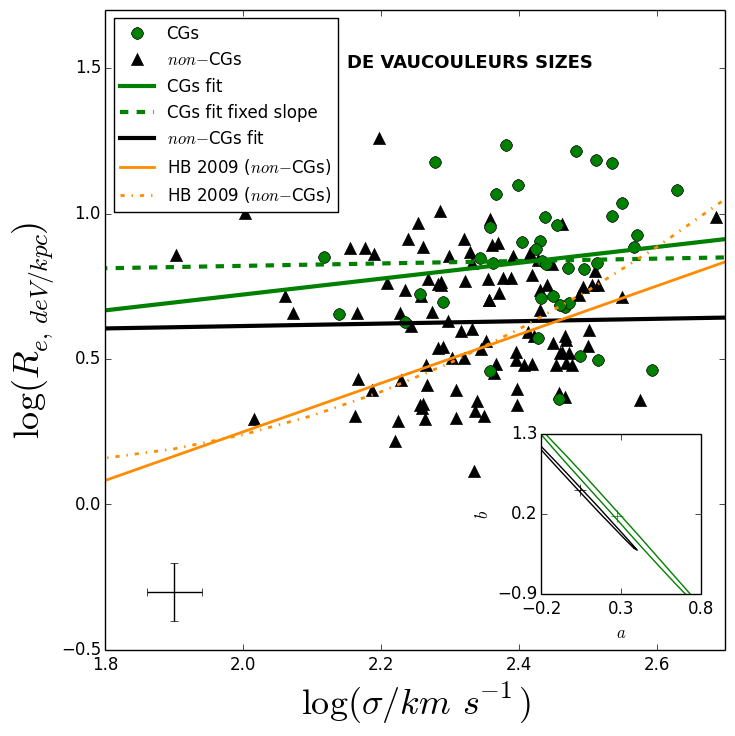}
\includegraphics[scale=0.45]{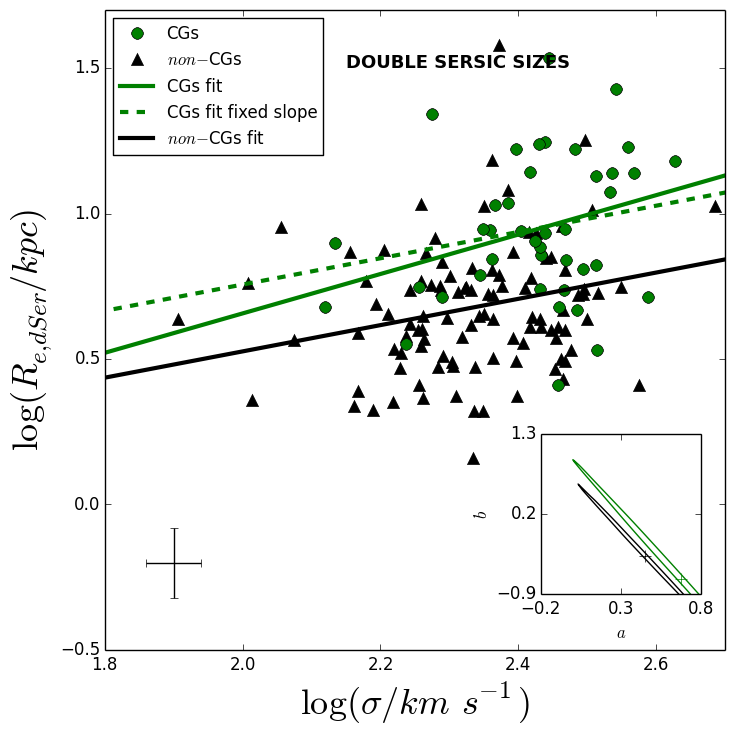}
\caption{$R_e- \sigma$ relation for \Bs (green circles) and \nBs (black triangles) for the spectroscopic COSMOS sample. Lines and symbols are as Fig. \ref{re_mass}. 
\label{re_sigma}}
\end{figure*}

We start by investigating the relation between $R_e$ and $M_*$, as shown in Fig. \ref{re_mass}.  Here we plot  all the group \Cs, regardless of whether they have a measured velocity dispersion.  We emphasize that our goal here is a {\em relative} comparison between \Cs and \nCs and so we do not adopt any completeness corrections, since completeness effects impact both populations the same way.  

The top plot uses values for $R_e$ and $M_*$ that are derived from the \dev profile fits.  These single-component profiles are known to fit a majority of early-types over a range of mass and redshift well \citep[e.g.,][]{vanderwel08}.  As a result, and given our expectation that the majority of early-types are fit even better with multiple components, the \dev profile-based $R_e$ and $M_*$ estimates can be associated with an inner, ``primary'' component, which is mostly unaffected by the addition of an outer envelope.  The size-mass relation using these estimates suggests a slightly larger and somewhat asymmetric scatter for $all-$\Cs, with a possible tendency for modest outliers with larger $R_{e,deV}$ at fixed $M_{*,deV}$, especially towards higher masses.

The bottom-left plot in Figure \ref{re_mass}  explores the relation derived for the double \ser fits.  The assumption of this profile leads to greater sensitivity to any outer components.  When we construct the relation with $R_{e,dSer}$ and $M_{*,dSer}$, a more substantial fraction of $all-$\Cs than \nCs appear to deviate from the relation defined by the \nCs and, indeed, followed by the majority of $all-$\Cs.  Note that the strongest outliers have $M_{*,dSer}$ estimates 2--3 times larger than the masses estimated from the \dev profile, indicating the significant amount of additional stellar material captured by the 2-component fit.

Finally, the bottom-right panel compares $R_{e,dSer}$ to $M_{*,deV}$.  In a very rough way, this relation attempts to highlight the effect on total size of a potential outer component (as revealed by the double \ser fit) as a function of the mass of the primary ``inner'' component ($M_{*,deV}$).  We now see the largest deviation in the sizes of $all-$\Cs at fixed stellar mass.  A plausible interpretation of these results is that a fraction of COSMOS group \Cs ---with a range in primary component stellar masses---have acquired outer components, which leads to larger measured sizes, particularly for the more sensitive multi-component models, and significantly larger total stellar masses.  The majority of $all-$\Cs, however, remain on the mass-size relation defined by \nCs.

Recently, a bending in the size-luminosity relation has also been pointed out by \cite{bernardi14}. This work has shown that galaxies with  $M_\ast >2 \times 10^{11} M_\odot$ have larger sizes than expected from a simple linear relation between size and total luminosity.

We fit lines to these correlations using the technique proposed by \cite{kelly07}, which employs a Bayesian framework to avoid biases introduced by inappropriate choices for the prior distributions of the independent variables. We also account for the errors in both the dependent and independent variables and allow for the intrinsic scatter.  In addition, we have tested for differences in the relations of the two populations by fixing the slope for $all-$\Cs to be the same as \nCs and fitting the zero-points.  

Table \ref{tab_fits} reports our obtained fit parameters. When all the parameters are free, given the small dynamical range probed, fits are not well constrained, and all relations are in statistical agreement.  Even after fixing the slope, when the \dev profile is adopted to estimate sizes and masses (top and bottom-left panel), $all-$\Cs and \nCs are best fit with relations that are  in statistical agreement.  In contrast, when double \ser sizes are considered, $all-$\Cs and \nCs are no longer statistically compatible.  As mentioned above, some of the difference in size between group $all-$\Cs and \nCs might stem from the somewhat different $M_\star$ ranges spanned by the two samples. To test this, we consider only the range of overlap (\M$\sim$ 10.6-11.6) and perform 10,000 random draws of both populations.  With this reduced dynamic range, we find that the parameters of the $R_e-M_\star$ relation are in agreement within  $3\sigma-$errors between the two samples in $48.5\pm0.5\%$ of the extractions when the \dev profile is adopted, and  in $29.1\pm0.4\%$  of the extractions when the \dser profile is adopted.  Despite the fact that the majority of $all-$\Cs follow the \nC relation, this test reinforces the presence of fundamental differences in the $R_e-M_\star$ between the two populations, even over a fixed range in $M_*$.  The statistical significance of these differences is perhaps underestimated by inadequacies of the simple linear fit we have adopted.  It is not able to capture the behavior that can be seen visually in Figure \ref{re_mass}, namely a deviation from a linear form driven by a modest fraction of $all-$\Cs. 

We wish to isolate this fraction of \Cs with potentially large outer components.  
We use the $R_{e,dSer}$--$M_{*,deV}$ relation in the bottom-right panel of Figure \ref{re_mass} to identify those galaxies located at $>1\sigma$  from the average relation determined by combining $all-$\Cs and \nCs together.   A total of 23  $all-$\Cs are flagged as outliers, representing the 13\%  of the population.  In \S \ref{discussion} we will examine whether this population is peculiar in other ways.  We note that there are two \nCs that are offset at high mass as well.  We visually inspected the \nCs and found that they are really characterized by an outer envelope, probably the result of a recent merger. In one case it is due to the presence of a close companion. Therefore, outer components might associated with \nCs as well, or there might be a mis-identification of some \nCs which are actually \Cs.

In the case of the \dev quantities, we compare our results to those presented  in \cite{hb09}  and \cite{bernardi09}.  The former considered early-type galaxies in the local universe in all environments, while the latter explored a sample of cluster CGs  in the local universe. They both used the same size  profile and IMF as we do. \cite{hb09} give both a linear fit and a quadratic one, \cite{bernardi09} analyzed two slightly different cluster CG samples and probed a smaller mass range than we do (\M$\sim$11-12).  Here we report all of their results. We note that these relations have been computed using circularized sizes, so they are not directly comparable to ours. In any case they provide us with a baseline comparison to results from the local Universe.  We see that the $R_{e,deV}$--$M_{*,deV}$ relation in COSMOS is compatible with the \cite{hb09} fits for early-type galaxies. In contrast,  the fit determined by \cite{bernardi09} for local cluster CGs is steeper than what we find here. This suggests evolution in the scaling relation (see also \citealt{trujillo04b, mcintosh05}), which we explore in more detail below.

\subsubsection{Velocity dispersion- Stellar mass relation}

Next, we examine the relation between stellar mass and velocity dispersion, shown in Figure \ref{sigma_mass}.  Of the three scaling relation variables ($R_e$, $M_*$, and $\sigma$), the velocity dispersion is likely the least sensitive to material in potential outer components because it is a luminosity-weighted quantity and therefore dominated by the galaxy's center.  Furthermore, we have applied (minor) corrections to derive estimates for $\sigma$ within $\frac{R_e}{8}$, making it further representative of just the center.  Given the results of the previous section, which demonstrated the similarity between \Cs and \nCs in observables more sensitive to the inner regions, we would therefore expect the $\sigma$--$M_*$ relation for \Cs and \nCs to be nearly identical.

Restricting our analysis to those galaxies with a measured velocity dispersion, Figure \ref{sigma_mass} shows the reult agrees with our expectations.  The $\sigma$--$M_*$ relation is virtually independent of the profile adopted\footnote{The correction of the velocity dispersion to the standard central velocity dispersion depends only weakly on the effective radius (see \S\ref{sec:vd})} and so we show only the \dev case.  COSMOS \Cs and \nCs are characterized by very similar relations, with \Cs simply occupying the statistical extreme of the general population.

Not surprisingly, the parameters of a linear fit to the $\sigma-M_\star$ relation are in agreement within the errors for the two populations, both when slopes are free and when they are fixed. Parameters  are also compatible with the relation obtained by \cite{hb09}. Deviations might be observed at low masses, but the number of galaxies with \M$<$10.8 is small in our sample.  In contrast, the results for cluster CGs presented in \cite{bernardi09} are significantly different, indicating a potential flattening of the $\sigma-M_\ast$ relation for CGs with time that we explore further below.

\subsubsection{Size - Velocity dispersion relation}

Finally, we turn to a  comparison of sizes and velocity dispersions, a scaling relation that relates an observable that can be sensitive to an outer component ($R_e$) to one that is expected to be almost entirely defined by the inner regions ($\sigma$).  Two versions of this relation are shown in Figure \ref{re_sigma}.  The left panel uses $R_{e,deV}$ and the right panel, $R_{e,dSer}$.  In both cases, we see more scatter than in the previous relations, but also a much larger separation between group \Cs and \nCs. 
At any given velocity dispersion, \Cs are systematically larger than \nCs by a factor of $\sim$1.5.  When the \dser profile is adopted, differences between \Cs and \nCs are even more striking with some \C outliers significantly above the \nC population.  This result can be interpreted as a more extreme version of the bottom-right panel of the size-mass relation (Figure \ref{re_mass}) which compared $R_{e,dSer}$ to $M_{*,deV}$.  Here we expect $\sigma$ to be even less affected than $M_{*,deV}$ by light at large radii.  Galaxies with significant outer components should therefore be even more distinct in the $R_{e,dSer}$--$\sigma$ relation. 

\begin{figure}
\centering
\includegraphics[scale=0.35]{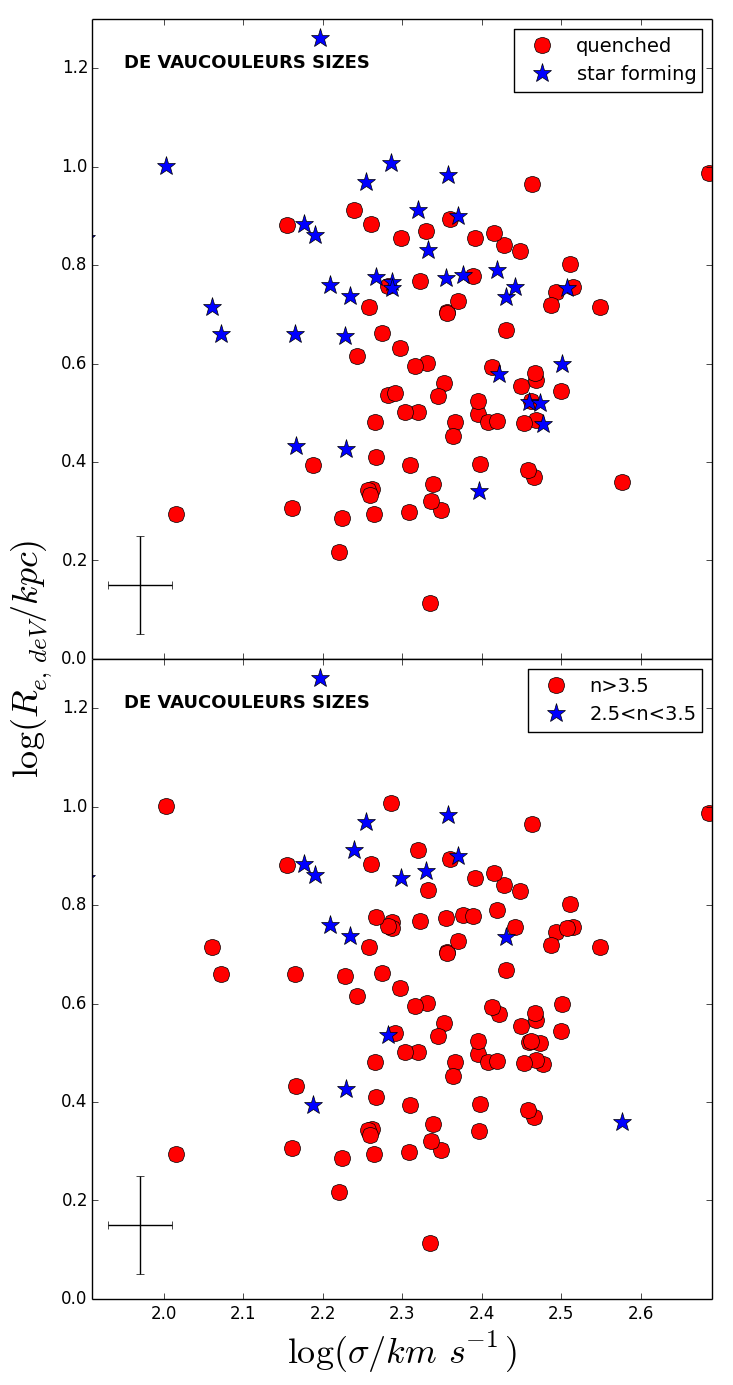}
\caption{$R_e- \sigma$ relation for \nBs, when  sizes  are measured adopting a \dev profile. Star forming and quenched (upper panel), high and low n (bottom panel) galaxies are shown separately. The typical errors are shown in the bottom left corner of each panel. 
\label{re_sigma_quenched}}
\end{figure}

We note that when the \dev profile is considered (left panel of Figure \ref{re_sigma}), the \nC sample shows a hint of two separate trends, only one of which appears consistent with \cite{hb09}. As shown in Fig. \ref{re_sigma_quenched}, this difference may be related to the star formation rate and morphology of our early-type sample.  We see that galaxies with somewhat lower \ser index ($2.5<n<3.5$), and ongoing star formation  have larger sizes than their counterparts with higher $n$, and less star formation, at any given $\sigma$.  This may be because   disks are typically more extended (larger) than ellipticals (e.g., \citealt{shen03}) and therefore when a disk is present, even along with a substantial bulge, the total estimated size could increase. 
These two trends are barely visible in the \C sample, likely because the \C sample is smaller and contains fewer disk galaxies.  When the sizes are derived from the double \ser fits, the two distinct trends in the \nC sample disappears.  This may be because the double \ser profile better accounts for the central bulge and disk in the more disk dominated galaxies at low $\sigma$ values, reducing their total sizes as compared to the single \dev profile.

\section{Comparisons to high- and low-redshift clusters}
In the previous section, we showed evidence that scaling relations for \Cs in COSMOS groups at $z\sim 0.6$  differ compared to the results of \cite{lauer07}  (who investigated a sample of 219 early-type galaxies which include the BCG sample described in \citealt{laine03}), and \cite{bernardi09} for clusters in the local universe. Setting aside systematic differences between the samples and measurement techniques, two physical factors may be at play: the much larger halo masses of clusters and potential redshift evolution. 

We now examine these possibilities  by studying the scaling relations obtained from two other samples: EDisCS, a sample of clusters that matches the COSMOS sample in redshift but includes much larger halos, and WINGS, a sample of clusters at $z \approx 0$.  The latter offers advantages over results from SDSS because it is more similar in both selection and measurements to our COSMOS sample which allows us to apply similar tests.

In what follows, we again stress that we can perform only {\em relative} comparisons, within each sample, between \Cs and \nCs. Systematics in sizes and masses, which are computed in different ways for different samples, as well as different definition of \Cs, prevent more quantiative and absolute comparisons.  In addition, as discussed below, the observations underpinning these samples have different surface brightness detection limits that, in principle, might affect the results.  

\subsection{Halo mass dependence}
\renewcommand{\tabcolsep}{2pt}
\begin{table*}
\centering
\caption{Scaling relations in clusters}
\begin{tabular}{cccccccc}
\hline
\hline
\multicolumn{7}{c}{EDisCS} \\
\multirow{2}{*}{Y} & \multirow{2}{*}{X}& \multirow{2}{*}{sample} & \multicolumn{3}{c}{free parameters} & \multicolumn{2}{c}{fixed slope}\\
				&						&						&Slope $a$  & Intercept $b$&Scatter &Slope $a$  & Intercept $b$\\ 
\hline
\multirow{2}{*}{$\sigma$} & \multirow{2}{*}{$M_{\star}$} & CGs &0.3$\pm$0.2&-1$\pm$3 & 0.005$\pm$0.003 &0.18 & 0.34$\pm$0.02\\
						&							&	{\it non-}CGs &0.18$\pm$0.03&0.3$\pm$0.3& 0.010$\pm$0.001&0.18 & 0.332$\pm$0.007\\
\multirow{2}{*}{$R_{e, deV+disk}$} & \multirow{2}{*}{$M_\star$} & CGs & 0.7$\pm$0.4 &-7$\pm$5 &0.02$\pm$0.01&0.70 & -6.88$\pm$0.03 \\
						&							&	{\it non-}CGs &0.70$\pm$0.08 &-7.0$\pm$0.9 &0.12$\pm$0.01&0.70 &-7.06$\pm$0.02\\
\multirow{2}{*}{$R_{e, deV+disk}$} & \multirow{2}{*}{$\sigma$} & CGs &1.0$\pm$0.5&-1$\pm$1 & 0.02$\pm$0.01& 0.6 & -0.23$\pm$0.03\\
						&							&	{\it non-}CGs &0.6$\pm$0.2 &-0.8$\pm$0.6 & 0.15$\pm$0.01& 0.6& -0.74$\pm$0.02\\
\hline
\hline
\multicolumn{7}{c}{WINGS} \\
\multirow{2}{*}{Y} & \multirow{2}{*}{X}& \multirow{2}{*}{sample} & \multicolumn{3}{c}{free parameters} & \multicolumn{2}{c}{fixed slope}\\
				&						&						&Slope $a$  & Intercept $b$&Scatter &Slope $a$  & Intercept $b$\\ 
\hline
\multirow{2}{*}{$\sigma$} & \multirow{2}{*}{$M_{\star, \, Ser}$} & CGs &0.5$\pm$0.3&-3$\pm$3 & 0.003$\pm$0.002& 0.38 & -1.97$\pm$0.02\\
						&							&	{\it non-}CGs &0.38$\pm$0.01&-1.8$\pm$0.1&0.0062$\pm$0.0004&  0.38 & -1.784$\pm$0.002\\
\multirow{2}{*}{$R_{e, ser}$} & \multirow{2}{*}{$M_{\star, \, Ser}$} & CGs & -- &-- &--& 0.68 & -6.41$\pm$0.05 \\
						&							&	{\it non-}CGs &0.68$\pm$0.02 &-6.7$\pm$0.3 &0.030$\pm$0.002& 0.68 &-6.701$\pm$0.005\\
\multirow{2}{*}{$R_{e, ser}$} & \multirow{2}{*}{$\sigma$} & CGs &1.0$\pm$0.4&-1$\pm$1 & 0.06$\pm$0.01& 0.59 & 0.04$\pm$0.03\\
						&							&	{\it non-}CGs &0.59$\pm$0.04 &-0.9$\pm$0.1 & 0.060$\pm$0.002& 0.59& -0.823$\pm$0.006\\
\hline
\end{tabular}
\tablecomments{Fits are of the form $\log Y=a\log X+ b$ with $M_\star$ 
in units of  $M_\odot$, $\sigma$ in units of $km\, s^{-1}$, $R_e$ in units of kpc. 
\label{tab_fits_cluster}}
\end{table*}

\begin{figure*}
\centering
\includegraphics[scale=0.3]{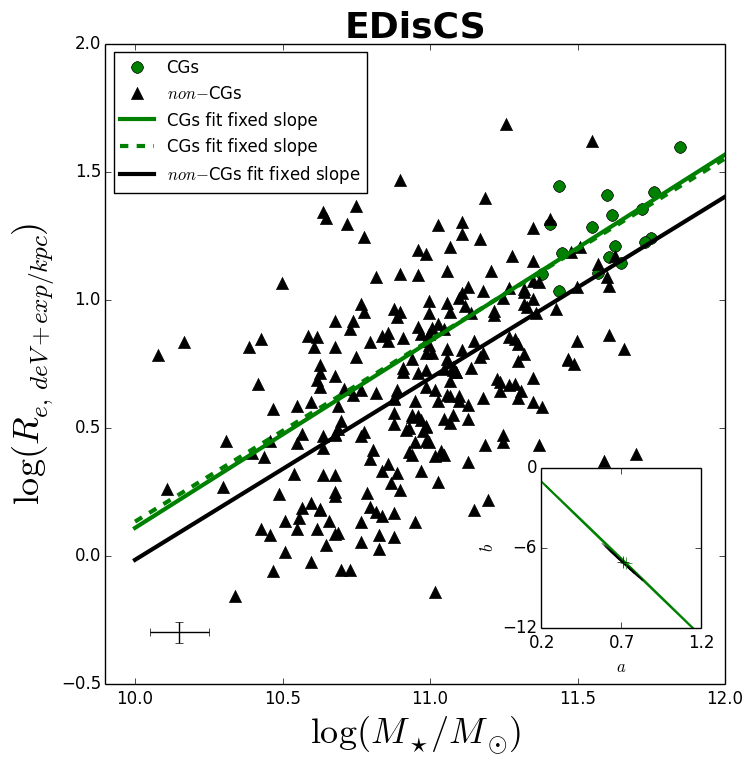}
\includegraphics[scale=0.3]{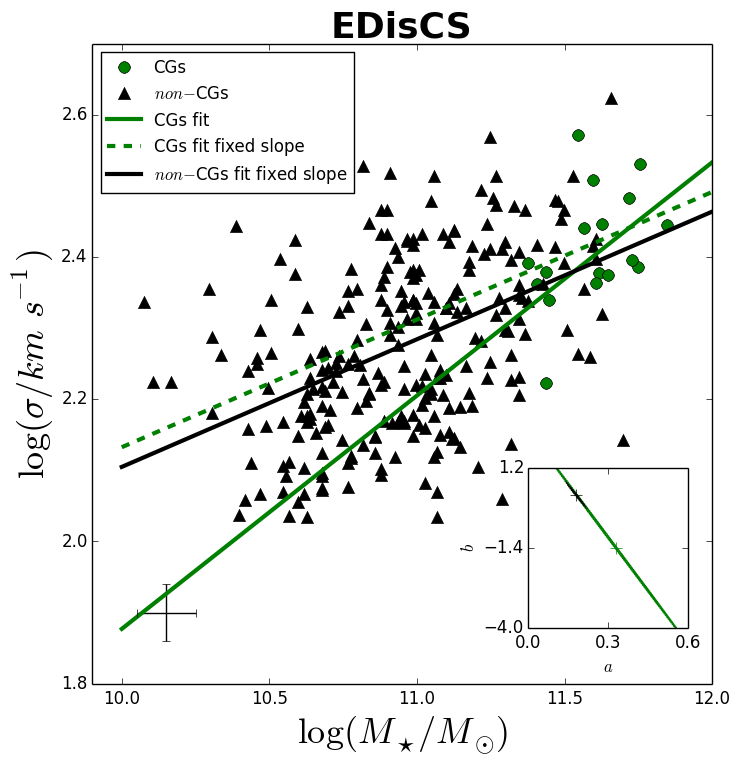}
\includegraphics[scale=0.3]{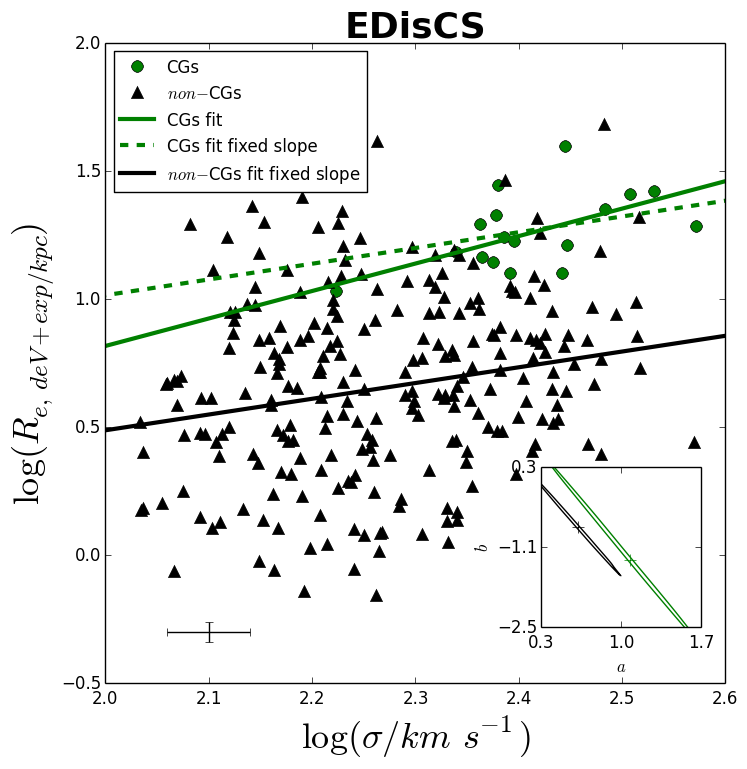}
\caption{Scaling relations for CGs (green circles) and {\it non-}CGs (black triangles) galaxies in clusters at $z\sim0.6$, drawn from the EDisCS clusters.   Solid lines are the linear fit to the relations, the green dotted line is the fit to the CGs when their slope is fixed to be the same as  {\it non-}CGs. The typical errors are shown in the bottom left corner of each panel.  In the insets, the 1-$\sigma$ contour errors to the fits are given. {\it Left panel}: $R_e - M_\star$ relation.  {\it Central  panel}: $\sigma - M_\star$ relation. {\it Right panel}: $R_e - \sigma$ relation. 
\label{ediscs}}
\end{figure*}

We first inspect the scaling relations obtained using the $z\sim0.6$ EDisCS sample.  It is important to remember that the COSMOS groups are representative of structures as massive as $M_{halo}\sim10^{14} \, M_\odot$, while EDisCS clusters extend up to $M_{halo}\sim 10^{15.2} \, M_\odot$ (see also \S\ref{discussion}).

Recall that for the EDisCS clusters we do not have detailed information on the profile fitting, so we cannot reference $M_*$ values to specific profile choices. We consider all galaxies in the sample, but when fitting EDisCS scaling relations, we take into account only galaxies above the mass completeness limit of \M$\sim 10.5$ \citep{vulcani11}.

The left panel of Figure \ref{ediscs} presents the size-mass relations for both \nCs and the \Cs in these $z\sim0.6$ clusters.  Combining {\it non-}CG  satellites and field galaxies together,\footnote{We have not detected any environmental dependence in the relations (in agreement with \citealt{maltby10, rettura10} at similar and higher redshift. See also Kelkar et al. submitted), so, to improve the statistics, we consider cluster satellites and field galaxies together.} few differences are detected compared to cluster CGs.  As in the COSMOS sample,  \Cs have  larger sizes for a given stellar mass (of a factor 1.2), but differences are marginally statistically significant (Tab.~\ref{tab_fits_cluster}).  We warn the reader that differences in size might be triggered by the fact that in the EDisCS sample \Cs and \nCs  have different mass distributions.  No \C population with extremely large sizes is apparent, but we caution that the EDisCS \C sample is small.

Moving to the $\sigma-M_\ast$ relation, the central panel in Fig. \ref{ediscs} shows that cluster CGs and \nCs have similar velocity dispersions. As we saw in COSMOS, we find no difference in the $\sigma-M_\ast$ relations for CGs in more massive halos. Linear fits are in agreement within the errors, both when all parameters are free and when a fixed slope is adopted (Tab.~\ref{tab_fits_cluster}).

However, again mirroring results from the COSMOS groups, we find that cluster CGs do stand out from {\it non-}CGs in the $R_e-\sigma$ relation (right panel in Fig. \ref{ediscs}): CGs are systematically offset high, by a factor of 3, and the parameters of the fits are significantly different. 

Despite the generally larger scatter in the EDisCS scaling relations (perhaps an indication that the EDisCS sample is less homogenous or subject to larger measurement uncertainties), the larger halo masses do not seem to result in a \C population that is more distinct from \nCs compared to what we find at group-mass scales.  If anything, there may be less evidence in this small sample for the type of outliers in size at fixed mass or $\sigma$ that seem characteristic of the COSMOS \C sample.

This is consistent with previous studies of the mass-size relation in the same COSMOS groups \citep{huertas13}, and as compared to high redshift clusters in \citep{delaye14}. These works also do not show a significant dependence of the mass-size relation on halo mass.
\subsection{Redshift dependence}

\begin{figure*}
\centering
\includegraphics[scale=0.3]{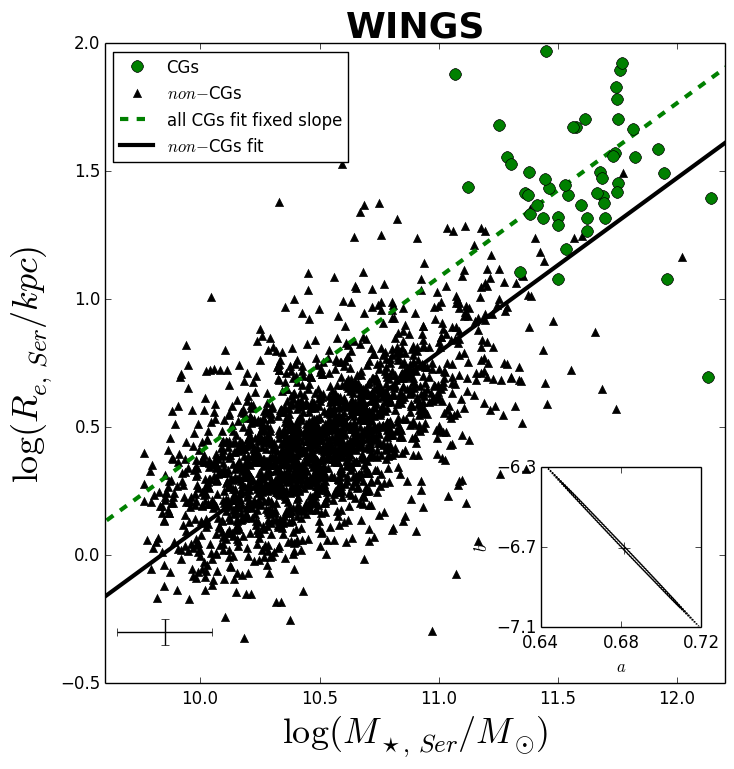}
\includegraphics[scale=0.3]{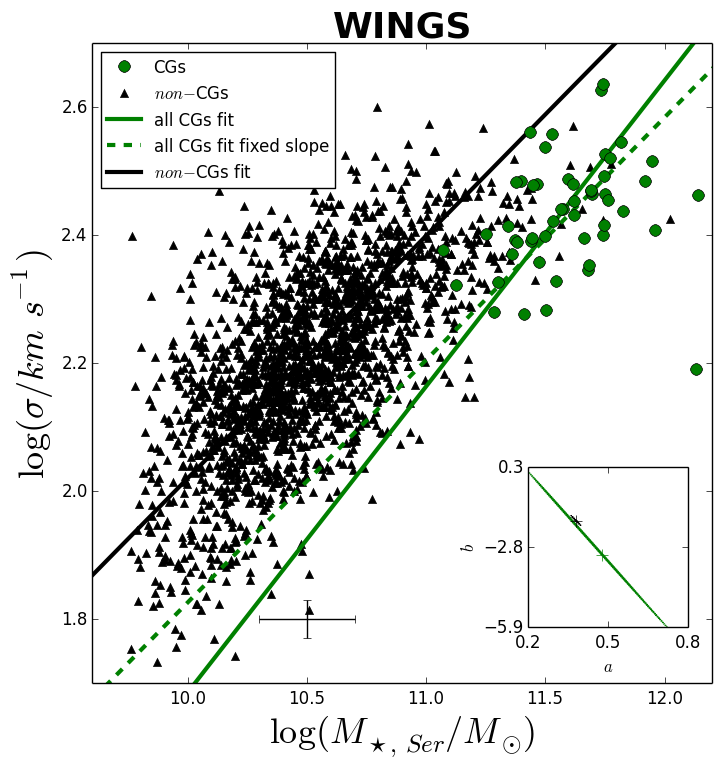}
\includegraphics[scale=0.3]{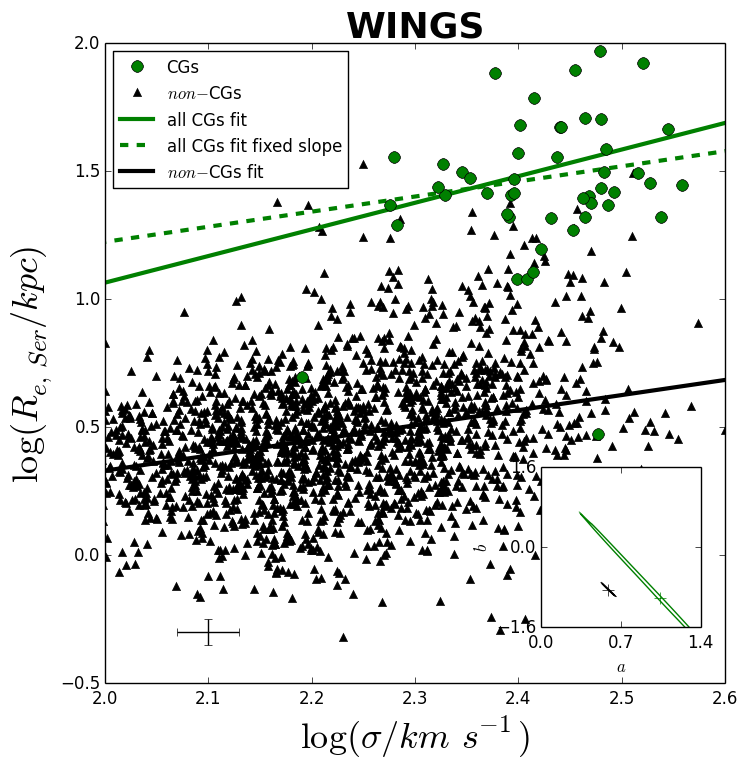}
\caption{Scaling relations for CGs (green circles) and {\it non-}CGs (black triangles) galaxies in clusters at $z\sim0$, drawn from the WINGS clusters. Solid lines are the linear fit to the relations, the green dotted line is the fit to the CGs when their slope is fixed to be the same as {\it non-}CGs. The typical errors are shown in the bottom left corner of each panel.  In the insets, the 1-$\sigma$ contour errors to the fits are given. {\it Left panel}: $R_e - M_\star$ relation.  {\it Central  panel}: $\sigma - M_\star$ relation.  {\it Right panel}: $R_e - \sigma$ relation. 
\label{wings}}
\end{figure*}

Having found little evidence for stronger offsets in \C properties as a function of halo mass, we now turn to the scaling relations for the $z$=0 cluster sample from WINGS (Figure \ref{wings}) to evaluate the potential impact of redshift evolution.

Recall that we  apply corrections to the stellar masses given in \cite{vulcani11} similar to those for the COSMOS galaxies, to reference the $M_*$ estimates to adopted single-\ser surface brightness profiles.   We  consider only galaxies above the mass completeness limit of \M$\sim$9.8.

Unlike the higher redshift EDisCS clusters, Fig. \ref{wings} shows how \Cs in $z = 0$ clusters strongly deviate from {\it non-}CGs in all scaling relations. At fixed velocity dispersion, for example, CGs have larger masses (by a factor of $\sim 3$) and larger sizes\footnote{We note that if we had not applied a mass correction relevant for the \ser profiles adopted, discrepancies between the two samples would be even larger.} (by a factor of 7) compared to \nCs (left and central panels of Fig. \ref{wings} respectively).  The consequence is a \C $R_e- \sigma$ relation that is completely offset with respect to the one for the {\it non-}CGs (right panel of Fig. \ref{wings}). The relations of the two populations are statistically different at high significance. We note that we could not find a meaningful fit for CGs in the $R_e-M_\star$ relation without fixing the slope. The parameters of the fits are given in Table \ref{tab_fits_cluster}.

Also different from the COSMOS or EDisCS samples at higher redshift, the $M_*$ distribution of WINGS \Cs is remarkably distinct: a gap between \Cs and \nCs is apparent (above all in the $R_e-M_\star$ relation), suggesting that galaxies slightly less massive than the \Cs have merged or been stripped.

These results recall those presented in \cite{valentinuzzi10}, who found that the mean size and mass of CGs have respectively increased by factors of $\sim$4 and $\sim$2 between z$\sim$0.6 (EDisCS) and z$\sim$0.04 (WINGS).

These findings reinforce the results of \cite{lauer07} and \cite{bernardi09}. Taken at face value, because the WINGS \Cs are substantially offset in all of the scaling relations, and because more information about the multi-component nature of the light profiles is not available, it is difficult to interpret Figure \ref{wings} in the context of structural changes affecting the ``primary'' inner regions of the \Cs versus the potential addition of outer regions.  However, as we discuss further in \S\ref{discussion}, the earlier snapshot provided by the COSMOS (and EDisCS) sample, strongly suggests the growth of outer components.  These components would increase the total $M_*$ and $R_e$ but leave $\sigma$ relatively unchanged, thus explaining the scaling relations observed at $z \approx 0$.   An important consideration that we examine below, however, is whether the appearance of these outer components at low redshift may be an observational effect owing to the deeper intrinsic luminosity densities  that can be probed in nearby studies.

\subsection{Surface brightness considerations for intra-sample comparisons}
In the previous sections we found that in clusters and groups at intermediate redshift, at a fixed stellar mass, CGs often have larger sizes than {\it non-}CGs, but very similar velocity dispersions. As a result, CGs and \nCs are particularly well-separated in the size-velocity dispersion plane.  In contrast, in local clusters, {\em all} \C scaling relations are offset compared to {\it non-}CGs.  Could the more extreme behavior at low-$z$ be caused by the fact that the low-density outskirts are easier to detect in nearby systems?

First, we stress that in the different samples, sizes have been measured adopting different profiles, hence quantitative intra-sample comparisons will likely be affected by systematic differences. In our COSMOS sample, size estimates obtained assuming  different profiles typically vary by 20\%. Therefore, we can only qualitatively contrast the observed trends. 

Next we make an appraisal of the imaging data used in the low-$z$ WINGS sample compared to the higher redshift EDisCS and COSMOS samples.  At $z \approx 0$, shallower observations should more easily detect material with the same physical projected  luminosity density.  For the different samples, we convert the reported surface brightness limits from $mag/arcsec^2$ to $L_\odot/kpc^2$.  In WINGS the surface brightness limit was computed setting  the detection threshold to 4.5$\sigma_{bg}/arcsec^2$ (see \S\ref{other_sample}).  This limit translates to a detection limit of $\mu_{Threshold}(V) \sim 25.7 mag/arcsec^2$ and corresponds to  $\sim 2.22\times 10^6 L_\odot/kpc^2$  at $z=0.05$.  In COSMOS the surface brightness limit is computed adopting a similar  detection threshold as in WINGS.
This limit translates to a detection limit of $\mu_{Threshold}(I) \sim 26.4\,  mag/arcsec^2$
and corresponds to $\sim 2.11\times 10^6 L_\odot/kpc^2$ and $\sim 5.82\times 10^6 L_\odot/kpc^2$ at $z=0.4$ and $z=0.8$ respectively. Given the similarity of EDisCS and COSMOS HST observations, we can adopt the same  thresholds for those two samples. This means that at least up to $z\sim0.6$ we would expect to detect potential outer envelopes down to the same physical projected density in all three data sets. 

We note that \cite{martizzi14} showed that a limit in surface brightness of $\mu(V)\sim25\, mag/arcsec^2$ (slightly shallower than that of the WINGS sample), is sufficient for detection of only 10-60\% of the toal CG mass, while a  2 magnitude deeper limit would allow for detection of 40-80\% of the total CG+envelope mass. Therefore, in all the datasets, we may be missing most of the mass related to the envelopes.  This entails that our measured sizes are actually lower  limits of the actual sizes of the objects. In addition, this is probably affecting more the lower-mass galaxies, therefore most likely  the  \nCs. 

Next, we consider a relative comparison. Assuming that the scale of the outskirts is similar in galaxies of all luminosities, we wish to test whether, for any galaxy, we are able to detect the same dynamic range in surface  luminosity density for all data sets. Using the mean sizes and luminosities for each sample, we find that the ratio of the peak average \C surface brightness compared to the limiting surface brightness is similar in the EDisCS and WINGS samples, while it is more than an order of magnitude smaller in COSMOS, owing to the fact that \Cs in the COSMOS groups are roughly an order of magnitude less luminous.  This means that low surface brightness components contributing the same {\em relative} fraction of luminosity to the primary component are intrinsically more difficult to detect in the COSMOS group sample.  Our utilization of single and 2-component profiles in \S\ref{res} helps overcome this limitation.

\section{Discussion}\label{discussion}
The main aim of this paper has been to carefully study the dynamical and structural properties of a sample of \C galaxies in groups at intermediate redshift.  By comparing \C properties to those of \nCs and tracking potential evolution, we seek to test theoretical explanations for the greater offsets for cluster \Cs at $z \approx 0$ than at $z \sim 0.6$.  We have focused on distinguishing between two types of mechanisms, both enhanced by the unique location of \Cs: 1) Violent processes (e.g., radial mergers) that drive deep structural changes \citep{boylan06} and 2) Smooth stellar accretion at large radii, as motivated by deep observations of the multi-component nature of early-type galaxies  (e.g., \citealt{dullo12} and references therein).  Both types of processes may be operating at lower levels in early-type galaxies generally, and offer different explanations for the more generic size growth observed since $z \sim 2$ (e.g.,  \citealt{trujillo06, van-dokkum08}).

Detecting and quantifying the size and mass of potential outer envelopes is a very delicate task.  Indeed, observations have to be deep enough to detect an excess of light in the profile at large distances from the center and often these regions are contaminated by the presence of other galaxies or confused with the sky. In addition, the choice of the profile to fit galaxies strongly influences the possibility of detecting such envelopes. Even when a double profile is adopted, it is not always easy to understand whether the outer component is representative of the external regions or of material not bounded to the galaxy.  

We have argued that the velocity dispersion is linked almost exclusively to the inner component of galaxies and not affected  by their outskirts.  Its value might be set early on, when the galaxy first assembled, and not influenced by events occurring in the outer parts. Size estimates track the inner component as well, but are also sensitive to an outer component.  While $\sigma$ does not depend on the choice of the profile, size does. As for masses, that scaled to the double \ser profile is an attempt to get information on the entire galaxy, while that  scaled to the \dev profile is a rough attempt to increase sensitivity to the inner component, alone.

In our analysis, we  detected differences between \Cs and \nCs in groups at intermediate redshift, in the scaling relations involving size, stellar mass, and velocity dispersion. Group \Cs at $z\sim0.6$ are systematically more massive with systematically higher velocity dispersions than their \nC counterparts. However, the $\sigma- M_\star$ relations are very similar. In contrast, scaling relations that involve the size of the galaxies are quite different for the two populations.  Even though  results depend on the profile adopted, for $M\geq10^{11}M_\odot$ there are clear signs for a sub-population of \Cs whose properties deviate from the normal trends.  Discrepancies are strongest when the size estimates we believe are most sensitive to outer components (double \ser) are plotted as a function of  $\sigma$, the observable most sensitive to the inner regions.  

Investigating galaxies in clusters at similar redshift, we obtain qualitatively similar results.\footnote{Once a common slope has been adopted, the $\sigma-M_\star$ relations obtained for EDisCS and COSMOS are in agreement within the uncertainties.}  We find that the discrepancy in $R_e$ between \Cs and \nCs is $\sim$1.5 times larger in clusters than in groups, despite the current lack of evidence for a significant population of cluster \C outliers at intermediate redshift, a possible selection effect.  The relative similarity between our group-scale results and those for clusters at similar redshifts, suggests that halo mass plays only a marginal role in influencing scaling relations, a topic we return to below.  Confirmation would require a profile analysis on a larger sample of intermediate-$z$ cluster \Cs with similar sensitivity to outer components as in our COSMOS analysis.

In local clusters, discrepancies are more pronounced: all scaling relations are different for CGs and {\it non-}CGs, suggesting that from $z\sim0.6$ to $z\sim0$  processes that impact either one or all galaxy properties are occurring in a way that affects size, mass and velocity dispersion differently.  In addition, there might be a dependence of mass and size growth on halo mass:  \cite{vulcani14} found that the relation between the stellar mass of the CG and the velocity dispersion of its host cluster (which is proportional to the halo mass) is steeper in WINGS than in EDisCS.  Similarly, we find hints that the size-halo mass relation is steeper in WINGS than in EDisCS. These findings suggest a faster evolution in both stellar mass and size for central galaxies located in more massive haloes.

Taken together, these results favor a scenario in which the primary component of the central galaxies (i.e. velocity dispersions) evolves little, but outer envelopes are accreted over time, and relatively rapidly from $z\sim0.6$ to $z\sim0$ (see also \citealt{valentinuzzi10}). 
From the similarity of the $\sigma- M_\star$ relation  for \Cs and \nCs at intermediate redshift, it seems that the early phase of formation of the two populations was  analogous, and that \Cs differentiate only at later epochs.  The new material accreted at large radius puffs up the galaxy without increasing its velocity dispersion and with only a modest increase in $M_*$.  However, as witnessed by the snapshot provided by the COSMOS sample, not all \Cs build these envelopes at the same rate.  We also note that \nCs in all samples show little evolution in dynamical scaling relations (see also \citealt{saglia10}), although there is intriguing evidence for a growing mass ``gap'' between satellite \nCs and the \C.

Before discussing the possible mechanisms responsible for the observed growth, we focus on  the properties of the group \Cs with larger sizes as a proxy for the presence of such envelopes and to see whether these \Cs are unique in terms of their halo and satellite populations.

\subsection{Indicators of candidate envelopes}
\begin{figure}
\centering
\includegraphics[scale=0.45]{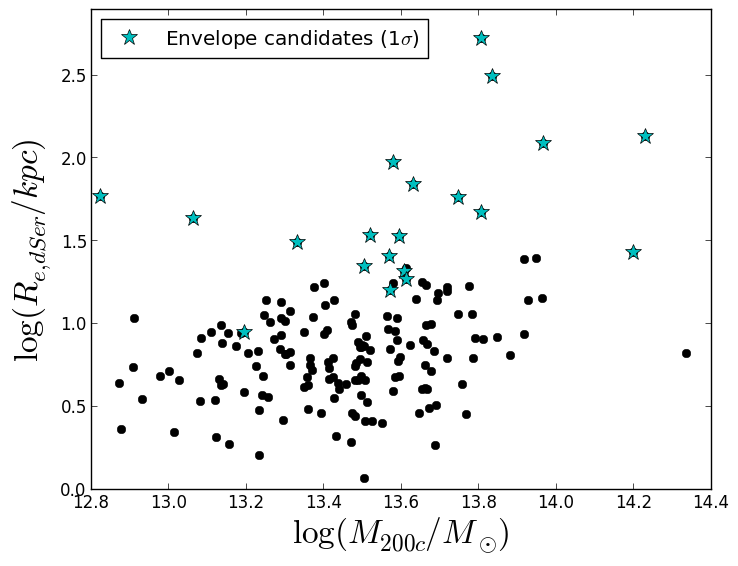}
\includegraphics[scale=0.45]{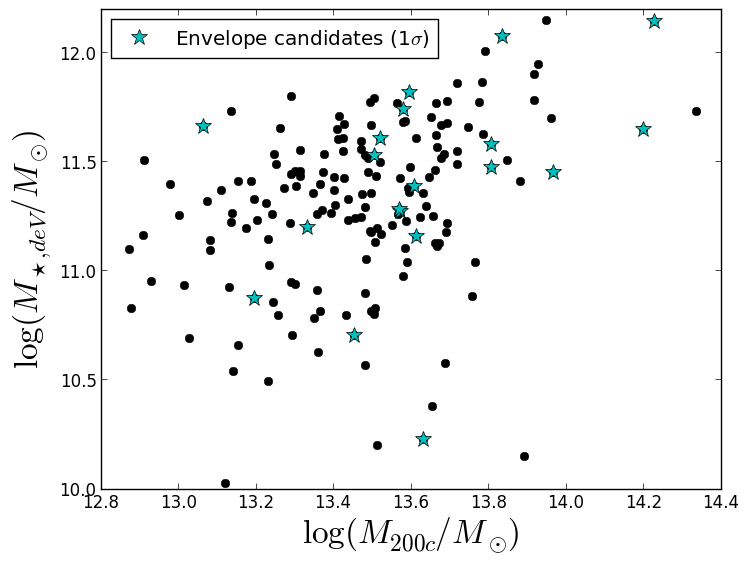}
\caption{$R_{e, dSer}- M_{200c}$ relation (upper panel) and $M_{\ast, deV}- M_{200c}$ relation (lower panel) for \Bs, when a double \ser profile is adopted.  
Galaxies located at $>$1$\sigma$ and  from the $R_e-M_\star$ relation (see \S \ref{relations} for details) are highlighted.  \label{re_Mhalo}}
\end{figure}

We select as outliers those galaxies that deviate by more than 1 standard deviation from the general $R_{e, dSer}-M_{\star, deV}$ relation when a \dser size  and \dev stellar mass are adopted (see Fig. \ref{re_mass}). With the hosts of candidate envelopes defined, we look for correlations with halo and satellite properties.

\begin{figure*}
\centering
\includegraphics[scale=0.4]{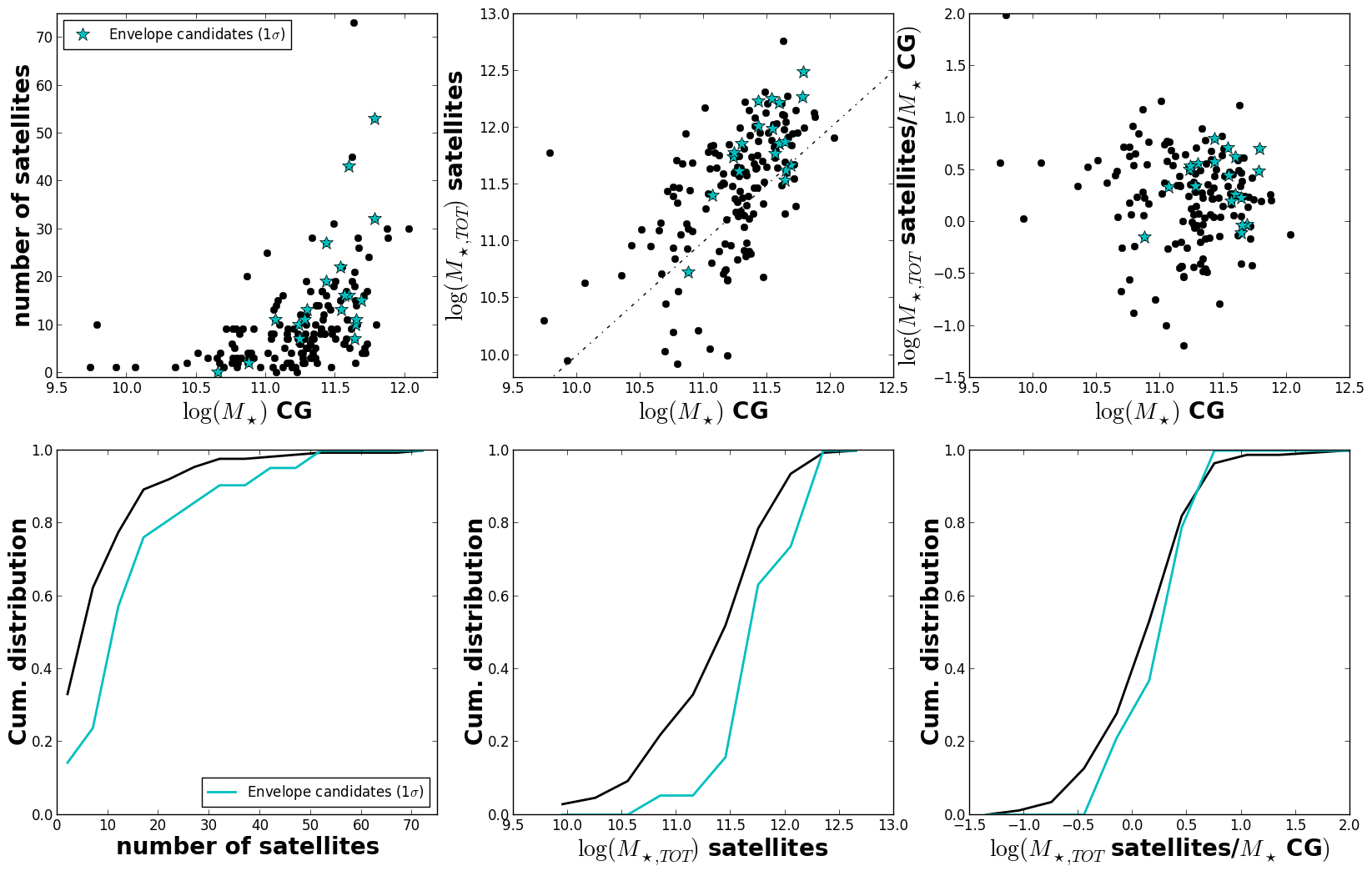}
\caption{Group satellite properties as a function of the \B mass. Upper panels: number of satellites (left), total mass in satellites (central) and total mass in satellites divided by the mass of the \B. In the central panel, dotted line indicates the 1:1 relation. Bottom panels: Cumulative distributions of the same quantities. Galaxies located at $>$1$\sigma$ from the $R_e-M_\star$ relation (see \S \ref{relations} for details) are highlighted. 
\label{satout}}
\end{figure*}

Figure \ref{re_Mhalo} shows the \dser $R_e- M_{200c}$ (upper panel) and the \dev $M_\ast- M_{200c}$ (lower panel) relations for all \Cs in the COSMOS group sample. \Cs with larger sizes, and hence most likely to have envelopes, are preferentially found in the most massive haloes ($\log M_{200c}>13.4$). However, not all massive haloes host CGs characterized by a large $R_e$. In the bottom panel, we see that \Cs with large envelopes are distributed across all $M_*$ as a function of $M_{200c}$.  As before, \C 
with a possible presence of an outer component do not stand out.

There is a hint, however, that envelopes may be preferentially found around \Cs with less ongoing star formation.  Exploiting the quenching parameter described in \cite{bundy10} and in \S\ref{add_data}, we find that 80$\pm$10\% of \Cs with candidate envelopes are quenched, while only 67$\pm$6\% are quenched among \Cs that lie on the standard $R_e$--$M_*$ relation. Probably due to our small sample statistics, percentages are not statistically different, but there might be a hint that the presence of an envelope induces quenching.

We now investigate potential correlations between the presence of an envelope and the properties of associated satellites.  The upper panels of Figure \ref{satout} correlate the \C mass with several group properties, while the bottom panels show the cumulative distributions of the same quantities.  
The left-most panels show the number of satellites\footnote{Here we are considering only groups with a confident spectroscopic association, far from the field edges, not potentially merging and with more than 4 members and considering only galaxies whose probability of being a group member is $P_{MEM}>$0.5 (see \citealt{george11, george12} for details).} more massive than \M=9.8\footnote{This is the COSMOS mass completeness limit as given in \cite{george12}.} as a function of the \C stellar mass.  Not surprisingly, the number of satellites increases with the stellar mass of the group's \C.  However,  despite the low statistics and significant scatter, we see that those \Cs with candidate envelopes tend to have more satellites at fixed \C $M_*$.  This is borne out in the cumulative distributions shown below (bottom-left).  The \Cs with envelopes have a satellite distribution that is noticeably shifted towards greater numbers.  The K-S test supports our finding, assessing that the distributions are different at the 99.5\% level.

The next set of panels examines the total stellar mass in satellites above the mass completeness limit with respect to the \C stellar mass.  More massive \Cs have systematically more massive satellites, but as previewed in the previous result, the population of \C envelope candidates tends to be associated with more satellite mass as well.  The K-S test statistically supports these result, claiming distributions are different at $>95\%$ level.

Some of the differences in the cumulative distributions in the first two columns could be driven by the fact discussed above that envelope candidates tend to be found among the most massive \Cs (and in the most massive halos).  To account for this effect we finally consider the ratio of mass in satellites to the mass of the \C in the right-most column of Figure \ref{satout}.  The distinction of \Cs with candidate envelopes is less striking, but still apparent,  even though it is not confirmed by the K-S test.  We note that we also looked for similar correlations with the fraction of quenched satellites (plots not shown), but did not find convincing trends.

Clearly larger samples are required to draw definitive conclusions, but our small COSMOS sample shows tantalizing hints of a correlation between the number and  total mass of satellite galaxies and the presence of outer components associated with the \C.  If these outer components are real and built through the stripping and accretion of satellites, it is plausible that \Cs living in halos with an overabundance of satellites are most likely to acquire such components.

In agreement with this scenario, a recent pilot study on one local CGs (NGC 3311, \citealt{coccato11}), showed that its outer envelope is currently accreting stars from the surrounding dwarf galaxies, which are being disrupted and accreted into the galaxy halo. The stellar population of the outer halo of NGC 3311 is indeed composed by a fraction of stars (<30\%), which are consistent with coming from dwarf galaxies.

\subsection{Origin and evolution of central galaxy envelopes}

\begin{figure}
\centering
\includegraphics[scale=0.45]{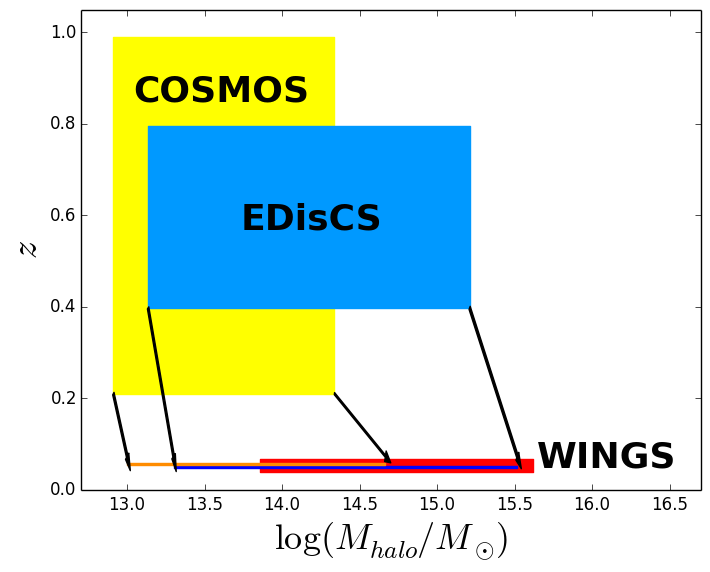}
\caption{Mass halo evolution of COSMOS groups and EDisCS cluster form $z\sim0.6$ to $z\sim$0, compared to the WINGS mass range. 
\label{evol_halo}}
\end{figure}

Our COSMOS results combined with low-$z$ cluster samples suggest that the velocity dispersions of early-type galaxies evolve little with time for both \Cs and \nCs. 
In contrast, at a fixed $\sigma$, the ratio of the  \C size to the \nC size changes from a factor of 2-3 at $z\sim0.6$  to a factor of 7 in the local universe. The ratio of masses similarly changes from a factor of $\sim1$ at $z\sim0.6$ to a factor of 3 in the local universe. 
When interpreting such trends, it is important to consider the progenitor-descendant relationship between the samples.  A schematic description is given in Figure \ref{evol_halo}.  From $z\sim0.6$ to $z=0$, both galaxies and haloes evolve. Using the halo mass growth rate computed by  \cite{fakhouri10}, EDisCS structures will evolve into the WINGS ones, hence their CGs will turn into the local universe CGs. In contrast, only the most massive COSMOS groups might become WINGS-like clusters at $z$=0 (Fig. \ref{evol_halo}). 

Theoretical arguments and other observational evidence suggest that this amount of growth is plausible.  Exploiting a semi-analytic model applied to the Millennium Simulation \citep{springel05}, \cite{delucia07} assess that central galaxies in clusters are expected to assemble half of their final stellar mass in the last $\sim 5-6$ Gyr.  Similarly, the change in size with time for \Cs is consistent with results from the literature that show that the evolution of galaxy sizes for spheroids with $n >$2.5 can be characterized by a power law of the form $R_e \propto (1 + z)^{-1.48 \pm 0.04}$ (see \citealt{conselice14} for a recent review).

Our results support a scenario in which \C size growth is occurring in the outer parts of galaxies, with the central parts in place at early times (see also, e.g., \citealt{carrasco10, vd10}). This indicates that the build up of massive galaxies is an inside-out process (e.g., \citealt{hopkins09}). While this scenario may apply to both central and satellites galaxies, our results suggest it is enhanced in CGs in all environments.  The massive \C at the center of the halo's potential well, may not only have a higher probability to be a site for interactions, but surely exerts greater tidal forces, which can distort nearby companions and bring galaxies down to the bottom of the potential well. 

Mergers are a primary channel for high-mass galaxy growth, and minor mergers in particular have been invoked to explain size evolution (e.g. \citealt{naab09, shankar13, shankar14}.  They can also induce  low levels of star formation in early-types at z$\sim$0.8, as well as add significant amount of stellar mass to these galaxies \citep{kaviraj11, kaviraj09} (but see \citealt{sonnenfeld14} who suggest that the outer regions of massive early types grow by accretion of stars and dark matter, while small amounts of dissipation and nuclear star formation conspire to keep the mass density profile constant and approximately isothermal). However, there is currently some controversy over whether the observed minor merger rate at $z<1$ is high enough to provide the required increase in sizes, with the most massive galaxies with $M_\star > 10^{11} M_\odot$ appearing to have enough minor mergers (e.g., \citealt{kaviraj09}) to produce this size evolution \citep{bluck12, poggianti13}, while this may not be the case for lower mass systems (e.g., \citealt{newman12}).

Some further insight may be gained by comparing our results to evolutionary predictions from the model described in \cite{nipoti12}. 
They estimated the merger-driven redshift evolution of the scaling relations considered here for typical massive early-type galaxies. They neglected dissipative processes, and thus maximized the evolution in surface density, under the assumption that the accreted satellites are spheroids. 

We stress that this comparison is a simplification in many aspects: indeed, the prescription gives an average growth, without taking into account stochastic events that can make each galaxy grow differently. Moreover, it adopts one-to-one mapping between halo and stellar mass with no scatter. 

With these caveats in mind, we use the expected mass evolution of halos from $z\sim0.6$, to $z\sim0$  to infer the evolution of the stellar mass for each CG in the COSMOS and EDisCS samples in the same redshift interval, using the $M_{halo}-M_\ast$ relation at $z\sim0.3$ and $z\sim0.5$ found by \cite{leauthaud12}. We then obtain the expected evolution in size and velocity dispersion given the evolution in stellar mass. Comparing the scaling relations obtained by evolving \C properties to the scaling relations obtained for the WINGS cluster \Cs we find that, in all the cases, the predicted evolution reduces the discrepancies between the high and low-$z$ samples:  the {\em relative} increase is greater for the sizes than for masses, and velocity dispersions show only a small decrease.  Very broadly speaking, the formalism of \cite{nipoti12} yields predictions that are consistent with the evolution inferred here.  A more quantitative comparison with an improved version of this model will be the subject of future work.  It will also be valuable to compare results from hydrodynamic simulations \citep[e.g.,][]{feldmann10, naab13}.

\section{Conclusions}
Utilizing followup VLT spectroscopy of COSMOS X-ray groups \citep{george11, george12}, we have studied the dynamical scaling relations of central and satellite galaxy populations in the previously unexplored regime of intermediate redshift ($z \sim 0.6$) and group-scale dark matter halos ($M_{\rm halo} \sim 10^{13} M_\odot$).  This regime is important for disentangling the role of mass and evolution among proposed mechanisms that attempt to explain offsets in the properties of $z \approx 0$ ``BCGs'' as compared to cluster and field early-type galaxies.

Our analysis and our main results can be summarized as follows:
\begin{itemize}
\item
Mindful of strong evidence for the multi-component nature of early-type surface-brightness profiles, we construct $R_e$ and $M_*$ estimators that differ in their sensitivity to light at large radii (adopting a \dev and double \ser profiles).

\item 
We use these $R_e$ and $M_*$ estimators in combination with $\sigma$ to examine scaling relations that reveal evidence for outer envelope components preferentially found among roughly $\frac{1}{7}$ of the \Cs (central galaxies) in our COSMOS sample.  

\item
We find that the \Cs with candidate envelopes tend to have larger $M_*$ and live in more massive halos.  We also find tantalizing evidence that the presence of an envelope is correlated with an overabundance of satellites at fixed \C $M_*$ and is more likely around quenched \Cs, although larger samples are needed to confirm these conclusions.

\item
We perform a relative comparison of the scaling relations of \Cs and \nCs in clusters at similar redshifts (from EDisCS) and at $z \approx 0$ (from WINGS).  The smaller sample of EDisCS \Cs shows a somewhat larger average offset in $R_e$ compared to \nCs, but no evidence for the more extreme \C outliers found in COSMOS, perhaps because the applied size estimator is not as sensitive to outer components.  Thus, the dependence of \C properties on $M_{\rm halo}$ at fixed redshift appears to be mild.  The \Cs in WINGS clusters, on the other  hand, are more drastically offset in all scaling relations, in agreement with other samples and work from SDSS.  
\end{itemize}

Taken together, these results favor a scenario in which the \Cs in groups and clusters rapidly accrete outer components, a process that is already underway by $z \sim 0.6$ (see also \citealt{valentinuzzi10}).  Assuming this process eventually effects all massive \Cs, the increases in mass and size caused by this accreted material appear to be sufficient to explain the offsets in CG scaling relations by $z \approx 0$.  Meanwhile, the central velocity dispersion of \Cs appears to evolve mildly if at all.  In line with many other arguments \citep[e.g.,][]{hopkins10b,greene13, perez13}, this inside-out growth, in which a primary inner component forms early, with further material deposited in the outskirts at late times, may not require violent episodes of radial major mergers \citep{boylan06} that alter the \Cs' core structure.

It remains to be seen whether the rate of merging, tidal disruption, and accretion is sufficient to explain the inferred growth of envelopes \citep[e.g.,][]{newman12}, although our initial comparison to the prescription of  \cite{nipoti12} is promising.  It will also be important to quantify the evolutionary trends inferred here with large, uniform samples that span larger ranges in both $M_{\rm halo}$ and redshift.  Finally, a key limitation is the inability to obtain robust multi-component fits that correspond to true physical components.  An accurate estimator of the mass in the proposed envelope component would be very valuable, although this challenge is surely coupled to our understanding of the ICL.   Recently, \cite{montes14} showed that the origin of the ICL is similar to that of the outskirts of galaxies. The two components can only be distinguish dynamically (e.g. \citealt{longobardi13}). 
In the future, spectroscopic measures may provide a means forward if outer components can be distinguished chemically \citep[e.g.,][]{greene13}.  Large IFU surveys like the upcoming MaNGA (Mapping Nearby Galaxies at Apache Point Observatory, Bundy et al.~2014) might then enable more robust decompositions that could be used to improve profile-fitting techniques.

\begin{acknowledgements}
We acknowledge the referee for her/his useful comments. We thank A. Moretti and the WINGS team for providing us their data. This work was supported by the World Premier International Research Center Initiative (WPI), MEXT, Japan. 
BV was also supported  by the
Kakenhi Grant-in-Aid for Young Scientists (B)(26870140) from the Japan Society for the Promotion of Science (JSPS). TT gratefully acknowledges support from the Packard Fellowship in the form of a Packard Fellowship.
\end{acknowledgements}

\bibliographystyle{apj}
\bibliography{biblio_bcg}

\end{document}